\documentclass[reqno,11pt]{amsart}
\usepackage{amscd,amssymb,verbatim}

\numberwithin{equation}{section}
\theoremstyle{plain}
\newtheorem{LinThm}[subsection]{Linearization Theorem}

\newcommand\alp{\alpha}         
\newcommand\bet{\beta}
\newcommand\gam{\gamma}         \newcommand\Gam{\Gamma}
         
\newcommand\eps{\varepsilon}

\newcommand\iot{\iota}

\newcommand\lam{\lambda}                \newcommand\Lam{\Lambda}
\newcommand\sig{\sigma}         \newcommand\Sig{\Sigma}

\newcommand\ome{\omega}         

\newcommand\calE{{\mathcal{E}}}

\newcommand\calH{{\mathcal{H}}}

\newcommand\calL{{\mathcal{L}}}

\newcommand\calO{{\mathcal{O}}}

\newcommand\calW{{\mathcal{W}}}

\newcommand\bfa{{\mathbf a}}

            \newcommand\bfD{{\mathbf D}}

\newcommand\RR{\mathbb{R}}

\newcommand\ZZ{\mathbb{Z}}

\newcommand\CC{\mathbb{C}}

 \newcommand\grg{{\mathfrak{g}}}

\newcommand\nek{,\ldots,}
\newcommand\sdp{\times \hskip -0.3em {\raise 0.3ex
\hbox{$\scriptscriptstyle |$}}} 

\newcommand\Dom{\operatorname{Dom}}

\newcommand\End{\operatorname{End\,}}

\newcommand\grad{\mathop {\rm grad}}
\newcommand\Hom{\operatorname {Hom}}
\newcommand\Id{\operatorname {Id}}

\newcommand\ind{\operatorname{ind}}

\newcommand\Irr{\operatorname{Irr\, }}

\newcommand\Ker{\operatorname{Ker}}

\newcommand\Coker{\operatorname{Coker}}
\newcommand\Lie{\operatorname{Lie}}

\newcommand\rk{\operatorname{rk}}

\newcommand\supp{\operatorname{supp}}

\newcommand\Tr{\operatorname{Tr}}

\newcommand\oc{{\bar{c}}}

\renewcommand\oe{{\bar{e}}}
\newcommand\of{{\bar{f}}}

\newcommand\oj{{\bar{j}}}
\newcommand\oJ{{\bar{J}}}

\newcommand\oM{{\bar{M}}}

\newcommand\hatA{{\hat{A}}}

\newcommand\hatsig{{\hat{\sig}}}

\newcommand\tilc{{\tilde{c}}}

\newcommand\tilD{{\tilde{D}}}
\newcommand\tilE{{\tilde{E}}}

\newcommand\tilU{{\tilde{U}}}
\newcommand\tilv{{\tilde{v}}}

\newcommand\tilW{{\tilde{W}}}

\newcommand\tilsig{{\tilde{\sig}}}

\renewcommand{\>}{\rangle}
\newcommand{\<}{\langle}

\theoremstyle{plain}
\newtheorem{Thm}[subsection]{Theorem}
\newtheorem{Cor}[subsection]{Corollary}
\newtheorem{Lem}[subsection]{Lemma}
\newtheorem{Prop}[subsection]{Proposition}
\newtheorem{Conjec}[subsection]{Conjecture}

\theoremstyle{definition}
\newtheorem{Def}[subsection]{Definition}

\theoremstyle{remark}

\newtheorem{Rem}[subsection]{Remark}

\newif\ifShowLabels
\ShowLabelstrue
\newdimen\theight
\def\TeXref#1{%
        \leavevmode\vadjust{\setbox0=\hbox{{\tt
                \quad\quad  {\small \textrm #1}}}%
        \theight=\ht0
        \advance\theight by \lineskip
        \kern -\theight \vbox to
        \theight{\rightline{\rlap{\box0}}%
        \vss}%
        }}%

\newcommand{\refs}[1]{Section ~\ref{S:#1}}
\newcommand{\refss}[1]{Subsection ~\ref{SS:#1}}
\newcommand{\reft}[1]{Theorem ~\ref{T:#1}}
\newcommand{\refl}[1]{Lemma ~\ref{L:#1}}
\newcommand{\refp}[1]{Proposition ~\ref{P:#1}}
\newcommand{\refc}[1]{Corollary ~\ref{C:#1}}

\newcommand{\refd}[1]{Definition ~\ref{D:#1}}
\newcommand{\refr}[1]{Remark ~\ref{R:#1}}
\newcommand{\refe}[1]{\eqref{E:#1}}

\newenvironment{thm}[1]%
        { \begin{Thm} \label{T:#1}  \ifShowLabels \TeXref{T:#1} \fi }%
        { \end{Thm} }

\renewcommand{\th}[1]{\begin{thm}{#1} \sl }
\renewcommand{\eth}{\end{thm} }

\newenvironment{lemma}[1]%
        { \begin{Lem} \label{L:#1}  \ifShowLabels \TeXref{L:#1} \fi }%
        { \end{Lem} }
\newcommand{\lem}[1]{\begin{lemma}{#1} \sl}
\newcommand{\elem}{\end{lemma}}

\newenvironment{propos}[1]%
        { \begin{Prop} \label{P:#1}  \ifShowLabels \TeXref{P:#1} \fi }%
        { \end{Prop} }
\newcommand{\prop}[1]{\begin{propos}{#1}\sl }
\newcommand{\eprop}{\end{propos}}

\newenvironment{corol}[1]%
        { \begin{Cor} \label{C:#1}  \ifShowLabels \TeXref{C:#1} \fi }%
        { \end{Cor} }
\newcommand{\cor}[1]{\begin{corol}{#1} \sl }
\newcommand{\ecor}{\end{corol}}

\newenvironment{conjec}[1]%
        { \begin{Conjec} \label{Conj:#1}  \ifShowLabels \TeXref{C:#1} \fi }%
        { \end{Conjec} }
\newcommand{\conj}[1]{\begin{conjec}{#1} \sl }
\newcommand{\econj}{\end{conjec}}

\newenvironment{defeni}[1]%
        { \begin{Def} \label{D:#1}  \ifShowLabels \TeXref{D:#1} \fi }%
        { \end{Def} }
\newcommand{\defe}[1]{\begin{defeni}{#1} \sl }
\newcommand{\edefe}{\end{defeni}}

\newenvironment{remark}[1]%
        { \begin{Rem} \label{R:#1}  \ifShowLabels \TeXref{R:#1} \fi }%
        { \end{Rem} }
\newcommand{\rem}[1]{\begin{remark}{#1}}
\newcommand{\erem}{\end{remark}}

\newcommand{\eq}[1]%
        { \ifShowLabels\newline \TeXref{E:#1} \fi
           \begin{equation} \label{E:#1} }
\newcommand{\eeq}{\end{equation}}

\newcommand{\prf}{ \begin{proof} }
\newcommand{\eprf}{ \end{proof} }
\newcommand{\Label}[1]{\label{#1}  \ifShowLabels \TeXref{#1} \fi }

\ShowLabelsfalse

\renewcommand{\d}{\text{\( \partial\)}}

\renewcommand{\b}{\bullet}

\newcommand{\n}{\nabla}

\newcommand{\E}{\calE}\newcommand{\W}{\calW}

\renewcommand{\L}{\calL}

\newcommand{\g}{{\Gam}}
\newcommand{\gc}{{\Gam(M,C(M))}}
\newcommand{\gme}{{\Gam(M,\E)}}
\newcommand{\ha}{^{1,0}}

\newcommand{\chern}{\operatorname{ch}\,}
\newcommand{\nLC}{\n^{\text{LC}}}
\newcommand{\wE}{\tilde\calE}

\renewcommand{\v}{\mathbf{v}}\renewcommand{\u}{\mathbf{u}}
\newcommand{\tv}{\mathbf{\tilv}}
\renewcommand{\i}{\sqrt{-1}\, }
\newcommand{\mD}{\bfD^{\text{mod}}}

\newcommand{\ve}{{\text{vert}}}
\newcommand{\h}{{\text{hor}}}

\newcommand{\an}{{\text{an}}}
\renewcommand{\top}{\text{top}}

\renewcommand{\H}{\bfD_a^2}

\newcommand{\ka}{K\"ahler }
\begin{document}
\title[Index theorem on non-compact manifolds]{Index theorem for equivariant Dirac operators
on non-compact manifolds
}
\author{Maxim Braverman}
\address{Department of Mathematics\\
        Northeastern University   \\
        Boston, MA 02115 \\
        USA
         }
\email{maxim@neu.edu}

\begin{abstract}
Let $D$ be  a (generalized) Dirac operator on a non-compact
complete Riemannian manifold $M$ acted on by a compact Lie group
$G$. Let $\v:M\to\grg=\Lie G$  be an equivariant map, such that
the corresponding vector field on $M$ does not vanish outside of a
compact subset. These data define an element of $K$-theory of the
transversal cotangent bundle to $M$. Hence, by embedding of $M$
into a compact manifold, one can define a topological index of the
pair $(D,\v)$ as an element of the completed ring of characters of
$G$.

We define an analytic index of $(D,\v)$ as an index space of
certain deformation of $D$ and we prove that the analytic and
topological indexes coincide.

As a main step of the proof, we show that index is an invariant of
a certain class of cobordisms, similar to the one considered by
Ginzburg, Guillemin and Karshon. In particular, this  means that
the topological index of Atiyah is also invariant under this class
of non-compact cobordisms.

As an application we extend the Atiyah-Segal-Singer equivariant
index theorem to our non-compact setting. In particular, we obtain
a new proof of this theorem for compact manifolds.
\end{abstract}
\maketitle \tableofcontents
\section{Introduction}\Label{S:introd}

Suppose $M$ is a complete Riemannian manifold, on which a compact
Lie group $G$ acts by isometries. To construct an index theory of
Dirac-type operators on $M$, one needs some additional structure
on $M$, which replaces the compactness. In this paper, this
additional structure is a $G$-equivariant map $\v:M\to\grg=\Lie
G$, such that the induced vector field $v$ on $M$ does not vanish
anywhere outside of a compact subset of $M$.  We call $\v$ a {\em
taming} map, and we refer to the pair $(M,\v)$ as a {\em tamed
$G$-manifold}.

Let $\E=\E^+\oplus\E^-$ be a $G$-equivariant $\ZZ_2$-graded
self-adjoint Clifford module over $M$. We refer to the pair
$(\E,\v)$ as a {\em tamed Clifford module}.

The pair $(\E,\v)$ defines an element in $K$-theory $K_G(T^*_GM)$
of transversal cotangent bundle, cf. \cite{AtiyahTE} and
\refss{topind} of this paper. Thus, using an embedding of (a
compact part of $M$) into a closed manifold and the excision
property (Th.~3.7 of \cite{AtiyahTE}), one can define an index of
$(\E,\v)$ as an element of the completed ring of characters of
$G$, cf. \refss{trsym}. We will refer to this index as the {\em
topological index} of the tamed Clifford module $(\E,\v)$ and we
will denote it by $\chi^{\top}_G(\E,\v)$. This index was
extensively studied by M.~Vergne \cite{Vergne96} and P.-E.~Paradan
\cite{Paradan1,Paradan2}.

The goal of this paper is to construct an analytic counterpart of
the topological index.

More precisely, we consider a Dirac operator
$D^\pm:L^2(M,\E^\pm)\to L^2(M,\E^\mp)$ associated to a Clifford
connection on $\E$ (here $L^2(M,\E)$ denotes the space of
square-integrable sections of $\E$). Let $f:M\to[0,\infty)$ be a
$G$-invariant function which increases fast enough at infinity
(see \refss{rescaling} for the precise condition on $f$). We
consider the {\em deformed Dirac operator} $D_{fv}=D+{\i}c(fv)$,
where $c:TM\simeq T^*M\to\End\E$ is the Clifford module structure
on $\E$. It turns out, cf. \reft{finite}, that {\em each
irreducible representation of $G$ appears in $\Ker D_{fv}$ with
finite multiplicities}. In other words, the kernel of the deformed
Dirac operator decomposes, as a Hilbert space, into (an infinite)
direct sum
\eq{Ker}
    \Ker D^\pm_{fv} \ = \ \sum_{V\in \Irr G}\, m^\pm_V\cdot V.
\end{equation}
Moreover, {\em the differences, $m_V^+-m_V^-$ are independent of
the choice of the function $f$ and the Clifford connection, used
in the definition of $D$}. Hence, these are invariants of the
tamed Clifford module $(\E,\v)$. We define the {\em analytic index
of $(\E,\v)$} by the formula
\[
    \chi^{\an}_G(\E,\v) \ := \
       \sum_{V\in \Irr G}\, (m^+_V-m^-_V)\cdot V.
\]

The main result of the paper is the {\em index
theorem~\ref{T:IndTh}}, which states that the analytic and
topological indexes coincide. The proof is based on an accurate
study of the properties of the analytic index. Some of these
properties will lead to new properties of the topological index
via our index theorem. More generally, the index formula allows us
to combine the analytic methods of this paper with the
$K$-theoretical methods developed by P.-E.~Paradan in
\cite{Paradan1,Paradan2}. Some simple examples are presented
bellow. For a more interesting application we refer the reader to
\cite{BrParadan}.

In \refs{cobord}, we introduce the notion of cobordism between
tamed Clifford modules. Roughly speaking, this is a usual
cobordism, which carries a taming map. Our notion of cobordism is
very close to the notion of non-compact cobordism developed by
V.~Ginzburg, V.~Guillemin and Y.~Karshon
\cite{GGK96,Karshon98,GGK-book}. We prove, that {\em the index is
preserved by a cobordism}. This result is the main technical tool
in this paper.

Suppose $\Sig\subset M$ is a compact $G$-invariant hypersurface,
such that the vector field $v$ does not vanish anywhere on $\Sig$.
We endow the open manifold $M\backslash{\Sig}$ with a complete
Riemannian metric and we denote by $(\E_\Sig,\v_\Sig)$ the induced
tamed Clifford module on $M\backslash{\Sig}$. In \refs{gluing}, we
prove that {\em  the tamed Clifford modules $(\E_\Sig,\v_\Sig)$
and $(\E,\v)$ are cobordant. In particular, they have the same
index}. We refer to this result as the {\em gluing formula}. Note,
that the gluing formula is a generalization of the excision
property for the index of transversally elliptic symbol, cf.
Th.~3.7 of \cite{AtiyahTE}.

It is worth noting that the gluing formula gives a non-trivial new
result even if $M$ is compact. In this case, it expresses the
usual equivariant index of $\E$ in terms of the index of a Dirac
operator on a non-compact, but, possibly, much simpler, manifold
$M_\Sig$.

The gluing formula takes especially nice form if $\Sig$ divides
$M$ into 2 disjoint manifolds $M_1$ and $M_2$. Let $(\E_1,\v_1)$
and $(\E_2,\v_2)$ be the restrictions of $(\E_\Sig,\v_\Sig)$ to
$M_1$ and $M_2$, respectively. Then the gluing formula implies
\[
    \chi^{\an}_G(\E,\v) \ = \ \chi^{\an}_G(\E_1,\v_1) \ +  \
    \chi^{\an}_G(\E_2,\v_2).
\]
In other words, {\em the index is additive}. This shows that the
index theory of non-compact manifolds is, in a sense, simpler than
that of compact manifolds (cf. \cite{Meinr-GS}, where a more
complicated gluing formula for compact manifolds is obtained).

In \refs{IndTh}, we use the gluing formula to prove that the
topological and analytical indexes of tamed Clifford modules
coincide. To this end we, first, consider a $G$-invariant open
relatively compact set $U\subset M$ with smooth boundary which
contains all the zeros of the vector field $v$. We endow $U$ with
a complete Riemannian metric and we denote by $(\E_U,\v_U)$ the
induced tamed Clifford module over $U$. As an easy consequence of
the gluing formula we obtain
\[
    \chi^{\an}_G(\E,\v) \ = \ \chi^{\an}_G(\E_U,\v_U).
\]
We then embed $U$ into a compact manifold $N$. By definition, cf.
\refss{topind}, the topological index $\chi_G^{\top}(\E,\v)$ is
equal to the index of a certain transversally elliptic operator
$P$ on $N$. In \refs{prIndTh} we give an explicit construction of
such an operator and by direct computations show that its index is
equal to $\chi^{\an}_G(\E_U,\v_U)$. We, thus, obtain the {\em
index formula}
\eq{indfor}
    \chi^{\an}_G(\E,\v) \ = \ \chi^{\top}_G(\E,\v).
\end{equation}

Atiyah, \cite{AtiyahTE}, showed that the kernel of a transversally
elliptic operator $P$ is a {\em trace class} representation of $G$
in the sense that $g\mapsto\Tr(g|_{\Ker P}), \ g\in G$ is well
defined as a distribution on $G$. It follows now from the index
formula \refe{indfor} that the index space of the operator
$D_{fv}=D+fc(v)$ is a (virtual) representation of trace class. In
other words the sum
\[
    T(g) \ = \ \sum_{V\in \Irr G}\, (m_V^+-m_V^-) \Tr(g|_V)
\]
converges to a distribution on $G$ (here $m_V^\pm$ are as in
\refe{Ker}). We don't know any direct analytic proof of this fact.
In particular, we don't know whether the individual sums
$\displaystyle\sum_{V\in\Irr{G}}m_V^\pm\Tr(g|_V)$ converge to
distributions on $G$.

As another application of the index formula, we see that the
topological index of Atiyah is invariant under our non-compact
cobordism. In particular, it satisfies the gluing formula. This
may be viewed as a generalization of the excision theorem~3.7 of
\cite{AtiyahTE}.

In \refs{ASS}, we consider the case when $G$ is a torus. Let
$F\subset M$ be the set of points fixed by the action of $G$.
Assume that the vector field $v$ does not vanish anywhere outside
of $F$. In \refss{linear}, we show that {\em  $(\E,\v)$ is
cobordant to a Clifford module over the normal bundle to $F$}.
This leads to {\em an extension of the Atiyah-Segal-Singer
equivariant index theorem to our non-compact setting}. As a
byproduct, we obtain a new proof of the classical
Atiyah-Segal-Singer theorem. This proof is an analytic analogue of
the proof given by Atiyah \cite[Lect.~6]{AtiyahTE}, Vergne
\cite[Part~II]{Vergne96} and Paradan \cite[\S4]{Paradan1}.

\subsection*{Acknowledgments}
This work started from a question of Yael Karshon and Victor
Guillemin. I would like to thank them for bringing my attention to
this problem and for valuable discussions.

I am very grateful to John Roe, who explained to me the modern
proofs of the cobordism invariance of the index on compact
manifolds. My proof of \reft{cobordinv} is based on the ideas I
have learned from John.

I am very thankful to Mich\`ele Vergne, for careful reading of the
original version of this manuscript, explaining me the connection
between tamed Clifford models and transversally elliptic symbols
and for bringing the works of Paradan to my attention.

\section{Index on non-compact manifolds}\Label{S:tamed}

In this section we introduce our main objects of study: tamed
non-compact manifolds, tamed Clifford modules, and the (analytic)
equivariant index of such modules.

\subsection{Clifford module and Dirac operator}\Label{SS:dirac}
First, we recall the basic properties of Clifford modules and
Dirac operators. When possible, we follow the notation of
\cite{BeGeVe}.

Suppose $(M,g^M)$ is a complete  Riemannian manifold.  Let $C(M)$
denote the Clifford bundle of $M$ (cf. \cite[\S3.3]{BeGeVe}),
i.e., a vector bundle, whose fiber at every point $x\in M$ is
isomorphic to the Clifford algebra $C(T^*_xM)$ of the cotangent
space.

Suppose $\E=\E^+\oplus\E^-$ is a $\ZZ_2$-graded complex vector
bundle on $M$ endowed with a graded action
\[
        (a,s) \ \mapsto \ c(a)s, \quad \mbox{where} \quad
                        a\in \gc, \ s\in \gme,
\]
of the bundle $C(M)$. We say that $\E$ is a {\em ($\ZZ_2$-graded
self-adjoint) Clifford module \/} on $M$ if it is equipped with a
Hermitian metric such that the operator $c(v):\E_x\to\E_x$ is
skew-adjoint, for all $x\in M$ and $v\in T_x^*M$.

A {\em Clifford connection}  on $\E$ is a Hermitian connection
$\n^\E$, which preserves the subbundles $\E^\pm$ and
\[
    [\n^\E_X,c(a)] \ = \ c(\nLC_X a), \quad
                \mbox{for any} \quad  a\in \gc, \ X\in\g(M,TM),
\]
where $\nLC_X$ is the Levi-Civita covariant derivative on $C(M)$
associated with the Riemannian metric on $M$.

The {\em Dirac operator \/} $D:\gme\to\gme$ associated to a
Clifford connection $\n^\E$ is defined by the following
composition
\[
  \begin{CD}
        \gme @>\n^\E>> \g(M,T^*M\otimes \E) @>c>> \gme.
  \end{CD}
\]
In local coordinates, this operator may be written as
$D=\sum\,c(dx^i)\,\n^\E_{\d_i}$. Note that $D$ sends even sections
to odd sections and vice versa: $D:\, \Gam(M,\E^\pm)\to
\Gam(M,\E^\mp)$.

Consider the $L^2$-scalar product on the space of sections $\gme$
defined by the Riemannian metric on $M$ and the Hermitian
structure on $\E$. By \cite[Proposition~3.44]{BeGeVe}, the Dirac
operator associated to a Clifford connection $\n^\E$ is formally
self-adjoint with respect to this scalar product. Moreover, it is
essentially self-adjoint with the initial domain smooth, compactly
supported sections, cf. \cite{Chernoff73},
\cite[Th.~1.17]{GromLaw83}.

\subsection{Group action. The index.}\Label{SS:G}
Suppose that a compact Lie group $G$ acts on $M$ by isometries.
Assume that there is given a lift of this action to $\E$, which
preserves the grading, the connection and the Hermitian metric on
$\E$. Then the Dirac operator $D$ commutes with the action of $G$.
Hence, $\Ker D$ is a $G$-invariant subspace of the space
$L^2(M,\E)$ of square-integrable sections of $\E$.

If $M$ is compact, then $\Ker D^\pm$ is finite dimensional. Hence,
it breaks into a finite sum \/ $\Ker D^\pm = \sum_{V\in \Irr G}\,
m^\pm_V\, V$, where the sum is taken over the set $\Irr G$ of all
irreducible representations of $G$. This allows one to defined the
{\em index}
\eq{char}
        \chi_G(D) \ = \ \sum_{V\in \Irr G}\, (m^+_V-m^-_V)\cdot V,
\end{equation}
as a virtual representation of $G$.

Unlike the numbers $m^\pm_V$, the differences $m^+_V-m^-_V$ do not
depend on the choice of the connection $\n^\E$ and the metric
$h^\E$. Hence, the index $\chi_G(D)$ depends only on $M$ and the
equivariant Clifford module $\E=\E^+\oplus\E^-$. We set
$\chi_G(\E):= \chi_G(D)$,
and refer to it as the {\em index} of $\E$.

\subsection{A tamed non-compact manifold} \Label{SS:tamed}
The main purpose of this paper is to define and study an analogue
of \refe{char} for a $G$-equivariant Clifford module over a
complete {\em non-compact \/} manifold. For this we need and
additional structure on $M$. This structure is given by an
equivariant map $\v:M\to\grg$, where $\grg$ denotes the Lie
algebra of $G$ and $G$ acts on it by the adjoint representation.
Note that such a map induces a vector field $v$ on $M$ defined by
\eq{v}
    v(x) \ := \ \frac{d}{dt}\Big|_{t=0}\, \exp{(t\v(x))}\cdot x.
\end{equation}
In the sequel, we will always denote maps to $\grg$ by bold
letters and the vector fields on $M$ induced by these maps by
ordinary letters.
\defe{tamed}
Let $M$ be a complete $G$-manifold. A {\em taming map} is a
$G$-equivariant map $\v:M\to\grg$, such that the vector field $v$
on $M$, defined by \refe{v}, does not vanish anywhere outside of a
compact subset of $M$. If $\v$ is a taming map, we refer to the
pair $(M,\v)$ as a {\em tamed $G$-manifold}.

If, in addition, $\E$ is a $G$-equivariant $\ZZ_2$-graded
self-adjoint Clifford module over $M$, we refer to the pair
$(\E,\v)$ as a {\em tamed Clifford module} over $M$.
\edefe
The index we are going to define depends on the (equivalence
class) of $\v$.

\subsection{A rescaling of $v$} \Label{SS:rescaling}
Our definition of the index uses certain rescaling of the vector
field  $v$. By this we mean the product $f(x)v(x)$, where
$f:M\to[0,\infty)$ is a smooth positive function. Roughly
speaking, we demand that $f(x)v(x)$ tends to infinity ``fast
enough" when $x$ tends to infinity. The precise conditions we
impose on $f$ are quite technical, cf. \refd{admissible}. Luckily,
our index turns out to be independent of the concrete choice of
$f$. It is important, however, to know that at least one
admissible function exists. This is guaranteed by \refl{rescaling}
bellow.

We need to introduce some additional notations.

For a vector $\u\in\grg$, we denote by $\L^\E_\u$ the
infinitesimal action of $\u$ on $\Gam(M,\E)$ induced by the action
on $G$ on $\E$. On the other side, we can consider the covariant
derivative $\n_{u}^\E:\Gam(M,\E)\to\Gam(M,\E)$ along the vector
field $u$ induced by $\u$. The difference between those two
operators is a bundle map, which we denote by
\eq{mu}
    \mu^\E(\u) \ := \ \n^\E_{u}-\L^\E_\u \ \in \ \End \E.
\end{equation}

We will use the same notation $|\cdot|$ for  the norms on the
bundles $TM, T^*M, \E$.  Let $\End(TM)$ and $\End(\E)$ denote the
bundles  of endomorphisms of $TM$ and $\E$, respectively. We will
denote by $\|\cdot\|$ the norms on these bundles  induced by
$|\cdot|$. To simplify the notation, set
\eq{nu}
    \nu=|\v|+\|\nLC v\|+\|\mu^\E(\v)\|+|v|+1.
\end{equation}
\defe{admissible}
We say that a smooth $G$-invariant function $f:M\to[0,\infty)$ on
a tamed $G$-manifold $(M,\v)$  is {\em admissible} for the triple
$(\E,\v,\n^\E)$ if
\eq{limv}
        \lim_{M\ni x\to\infty}\, \frac{f^2|v|^2}{
          |df||v|+f\nu+1 }  \ = \ \infty.
\end{equation}
\edefe
\lem{rescaling}
Let $(\E,\v)$ be a tamed Clifford module and let $\n^\E$ be a
$G$-invariant Clifford connection on $\E$. Then  there exists an
admissible function $f$ for the triple $(\E,\v,\n^\E)$.
\elem
We prove the lemma in \refs{prrescaling} as a particular case of a
more general \refl{admisb}.

\subsection{Index on non-compact manifolds}\Label{SS:noncomind}
We use the Riemannian metric on $M$, to identify the tangent and
the cotangent bundles to $M$. In particular, we consider $v$ as a
section of \/ $T^*M$.

Let $f$ be an admissible function. Consider the {\em deformed
Dirac operator}
\eq{Dv}
        D_{fv} \ = \ D \ + \ {\i}c(fv).
\end{equation}
This is again a $G$-invariant essentially self-adjoint operator on
$M$, cf. the remark on page 411 of \cite{Chernoff73}.

Our first result is the following
\th{finite}
Suppose $f$ is an admissible function.  Then

1. \ The kernel of the deformed Dirac operator $D_{fv}$
decomposes, as a Hilbert space, into an infinite direct sum
\eq{finite}
        \Ker D^\pm_{fv} \ = \ \sum_{V\in \Irr G}\, m^\pm_V\cdot V.
\end{equation}
In other words, each irreducible representation of $G$ appears in
$\Ker D^\pm_{fv}$ with finite multiplicity.

2. \ The differences $m^+_V-m^-_V$ $(V\in\Irr G)$ are independent
of the choices of the admissible function $f$ and the
$G$-invariant Clifford connection on $\E$, used in the definition
of $D$.
\eth
The proof of the first part of the theorem is given in
\refs{prfinite}. The second part of the theorem will be obtained
in  \refss{grch} as an immediate consequence of \reft{cobordinv}
about cobordism invariance of the index.

We will refer to the pair $(D,\v)$ as a {\em tamed Dirac
operator}. The above theorem allows to defined the index of a
tamed Dirac operator:
\[
        \chi_G(D,\v) \ := \ \chi_G(D_{fv})
\]
using \refe{char}. Note, however, that now the sum in the right
hand side of \refe{char} is infinite.

Since $\chi_G(D,\v)$ is independent of the choice of the
connection on $\E$, it is an invariant of the tamed Clifford
module $(\E,\v)$. This allows us to define  the {\em (analytic)
index of a tamed Clifford module} $(\E,\v)$ by
$\chi^{\an}_G(\E,\v):=\chi_G(D,\v)$, where $D$ is the Dirac
operator associated to some $G$-invariant Clifford connection on
$\E$.

Most of this paper is devoted to the study of the properties of
$\chi^{\an}_G(\E,\v)$. In \refs{cobord} we will show that it is
invariant under certain class of cobordisms. In particular, this
implies that $\chi^{\an}_G(\E,\v)$ depends only on the cobordism
class of the map $\v$. In some cases, one can give a very simple
topological description of the cobordism classes of $\v$. In the
next subsection, we do it for the most important for applications
case of {\em topologically tame manifolds}.

\subsection{Topologically tame manifolds}\Label{SS:toptame}
Recall that a (non-compact) manifold $M$ is called {\em
topologically tame} if it is diffeomorphic to the interior of a
compact manifold $\oM$ with boundary.

Suppose $M$ is a topologically tame manifold and let us fix a
diffeomorphism $\phi$ between $M$ and the interior of a compact
manifold with boundary $\oM$. A small neighborhood  $U$ of the
boundary $\d\oM$ of $\oM$ can be identified as
\eq{neigh}
    U \ \simeq \ \d\oM\times[0,1).
\end{equation}

Let $\v:M\to \grg$ be a taming map. Then it induces, via $\phi$
and \refe{neigh}, a map $\d\oM\times[0,1)\to\grg$, which we will
also denote by $\v$. Hence, for each $t\in[0,1)$, we have a map
$\v_t:\d\oM\to\grg$, obtained by restricting $\v$ to
$\d\oM\times\{t\}$. It follows from \refd{tamed}, that
$\v_t(x)\not=0$ for small $t$ and any $x\in\d\oM$. Thus
$\v_t(x)/\|\v_t(x)\|$ defines a map from $\d\oM$ to the unit
sphere $S_{\grg}$ in $\grg$. Clearly, the homotopy class of this
map does not depend on $t$, nor on the choice of the splitting
\refe{neigh}.  We denote by $\sig(\v)$ the obtained homotopy class
of maps $\d\oM\to S_{\grg}$. The following proposition is a direct
consequence of cobordism invariance of the index
(\reft{cobordinv}).
\prop{toptame}
If $M$ is a topologically tame manifold, then the index
$\chi^{\an}_G(\E,\v)$  does not change if we change the map
$\v:M\to\grg$, provided $\sig(\v)$ does not change.
\eprop

\section{Cobordism invariance of the index}\Label{S:cobord}

In this section we introduce the notion of cobordism between tamed
Clifford modules and tamed Dirac operators. We show that the
analytic index, introduced in \refss{noncomind}, is invariant
under a cobordism. This result will serve as a main technical tool
throughout the paper. In particular, we use it in the end of this
section to show that the index is independent of the choice of the
admissible function and the Clifford connection on $\E$.

\subsection{Cobordism between tamed $G$-manifolds}\Label{SS:cobordM}
Note, first, that, for cobordism to be meaningful, one must make
some compactness assumption. Otherwise, any manifold is cobordant
to the empty set via the noncompact cobordism $M\times[0,1)$.
Since our manifolds are non-compact themselves, we can not demand
cobordism to be compact. Instead, we demand the cobordism to carry
a taming map to $\grg$.
\defe{cobordM}
A {\em cobordism \/} between tamed $G$-manifolds $(M_1,\v_1)$ and
$(M_2,\v_2)$ is a triple $(W,\v, \phi)$, where
\begin{enumerate}
  \item  $W$ is a complete Riemannian $G$-manifold with boundary;
  \item  $\v:W\to\grg$ is a smooth $G$-invariant map, such that
  the corresponding vector field $v$  does
  not vanish anywhere outside of a compact subset of $W$;
  \item  $\phi$ is a $G$-equivariant, metric preserving
  diffeomorphism between a neighborhood $U$ of the boundary $\d W$
  of $W$ and the disjoint union $\big(M_1\times[0,\eps)\big)\,
  \bigsqcup\, \big(M_2\times(-\eps,0]\big)$. We will refer to $U$
  as the {\em neck} and we will identify it with $\big(M_1\times[0,\eps)\big)\,
  \bigsqcup\, \big(M_2\times(-\eps,0]\big)$.
  \item the restriction of $\v\big(\phi^{-1}(x,t))$ to
  $M_1\times[0,\eps)$ (resp. to $M_2\times(-\eps,0]$)
  is equal to $\v_1(x)$ (resp. to $\v_2(x)$).
\end{enumerate}
\edefe
\rem{Karshon}
A cobordism in the sense of \refd{cobordM} is also a cobordism in
the sense of Guillemin, Ginzburg and Karshon
\cite{GGK96,Karshon98,GGK-book}. If $G$ is a circle, one can take
$|fv|^2$ (where $f$ is an admissible function) as an abstract
moment map. It is not difficult to construct an abstract moment
map out of $\v$ in the general case.
\erem

\subsection{Cobordism between tamed Clifford modules}\Label{SS:cobordE}
We now discuss our main notion -- the cobordism between tamed
Clifford modules and tamed Dirac operators. Before giving the
precise definition let us fix some notation.

If $M$ is a Riemannian $G$-manifold, then, for any interval
$I\subset\RR$, the product $M\times I$ carries natural Riemannian
metric and $G$-action. Let $\pi:M\times I\to M, \ t:M\times{I}\to
I$ denote the natural projections. We refer to the pull-back
$\pi^*\E$ as a vector bundle {\em induced} by $\E$. We view $t$ as
a real valued function on $M$, and we denote by $dt$ its
differential.

\defe{cobordE}
Let  $(M_1,\v_1)$ and $(M_2,\v_2)$ be tamed $G$-manifolds. Suppose
that each $M_i$, $i=1,2$, is endowed with a $G$-equivariant
self-adjoint Clifford module $\E_i=\E^+_i\oplus\E^-_i$. A {\em
cobordism \/} between the tamed Clifford modules  $(\E_i,\v_i)$,
$i=1,2$, is a cobordism $(W,\v,\phi)$ between $(M_i,\v_i)$
together with a pair $(\E_W,\psi)$, where
\begin{enumerate}
  \item $\E_W$ is a $G$-equivariant (non-graded) self-adjoint
  Clifford module over $W$;
  \item $\psi$ is a $G$-equivariant isometric isomorphism between
  the restriction of $\E_W$ to $U$ and the Clifford module induced
  on the neck $\big(M_1\times[0,\eps)\big)\bigsqcup
  \big(M_2\times(-\eps,0]\big)$ by $\E_i$.
  \item On the neck $U$ we have
    \(\displaystyle
        c(dt)|_{\psi^{-1}\E_i^\pm} \ = \ \pm \i.
    \)
\end{enumerate}
\edefe
In the situation of \refd{cobordE}, we say that the tamed Clifford
modules $(\E_1,\v_1)$ and $(\E_2,\v_2)$  are {\em cobordant} and
we refer to $(\E_W,\v)$ as a cobordism between these modules.

\vskip0.2cm
\rem{M12-M}
Let $\E^{\text{op}}_1$ denote the Clifford module $\E_1$  with the
opposite grading, i.e., $\E^{\text{op}\pm}_1=\E_1^\mp$. Then,
$\chi^{\an}_G(\E_1,\v_1)=  -\chi^{\an}_G(\E^{\text{op}}_1,\v_1)$.

Consider the Clifford module $\E$ over the disjoint union
$M=M_1\sqcup{M_2}$ induced by the Clifford modules
$\E_1^{\text{op}}$ and $\E_2$. Let $\v:M\to\grg$ be the map such
that $\v|_{M_i}=\v_i$. A cobordism between $(\E_1,\v_1)$ and
$(\E_2,\v_2)$ may be viewed as  a cobordism between $(\E,\v)$ and
(the Clifford module over) the empty set.
\erem
\vskip0.2cm

One of the main results of this paper is the following theorem,
which asserts that the index is preserved by a cobordism.

\th{cobordinv}
   Suppose $(\E_1,\v_1)$ and $(\E_2,\v_2)$  are cobordant tamed Clifford
   modules. Let $D_1, D_2$ be Dirac operators associated to
   $G$-invariant Clifford connections on $\E_1$ and
   $\E_2$, respectively. Then, for any admissible functions
   $f_1, f_2$,
   \[
        \chi_G\big(\, D_1+{\i}c(f_1v_1)\, \big)
         \ = \ \chi_G\big(\, D_2+{\i}c(f_2v_2)\, \big).
   \]
\eth
The proof of the theorem is given in \refs{prcobordinv}. Here we
only explain the main ideas of the proof.

\subsection{The scheme of the proof}\Label{SS:sprcobordinv}
By \refr{M12-M}, it is enough to show that, if $(\E,\v)$ is
cobordant to (the Clifford module over) the empty set, then
$\chi_G(D_{fv})=0$ for any admissible function $f$.

Let $(W,\E_W,\v)$ be a cobordism between the empty set and
$(\E,\v)$ (slightly abusing the notation, we denote by the same
letter $\v$ the taming maps on $W$ and $M$).

In \refs{prrescaling} we define the notion of an admissible
function on a cobordism  $(W,\E_W,\v)$ analogous to
\refd{admissible}. Moreover, we show (cf. \refl{admisb}) that, if
$f$ is an admissible functions on $(M,\E,\v)$, then there exists
an admissible function on $(W,\E_W,\v)$, whose restriction to $M$
equals $f$. By a slight abuse of notation, we will denote this
function by the same letter $f$.

Let $\tilW$ be the manifold obtained from $W$ by attaching a
cylinder to the boundary, i.e.,
\[
    \tilW \ = \ W \, \sqcup \,
                \big(\, M\times(0,\infty)\, \big).
\]
The action of $G$, the Riemannian metric,  the map $\v$, the
function $f$ and the Clifford bundle $\E_W$ extend naturally from
$W$ to $\tilW$.

Consider the exterior algebra
$\Lam^\b\CC=\Lam^0\CC\oplus\Lam^1\CC$. It has two (anti)-commuting
actions $c_L$ and $c_R$ (left and right action) of the Clifford
algebra of $\RR$, cf. \refss{wE}. Define a grading of $\wE$ and a
Clifford action $\tilc:T^*\tilW\to\End\wE$   by the formulas
\[
   \wE^+ \ := \ \E_W\otimes\Lam^0; \quad  \wE^- \ := \
   \E_W\otimes\Lam^1; \qquad
   \tilc(v) \ := \ \i c(v)\otimes c_L(1) \quad (v\in T^*\tilW).
\]
Let $\tilD$ be a Dirac operator on $\wE$ and consider the operator
\footnote{Note that $v$ might vanish somewhere near
infinity on the cylindrical end of $\tilW$. In particular, the
index of $\tilD_{fv}$ is not defined in general.}
$\tilD_{fv}:=\tilD+c(fv)$.

Let $p:\tilW\to\RR$ be a map, whose restriction to
$M\times(1,\infty)$ is the projection on the second factor, and
such that $p(W)=0$. For any $a\in\RR$, consider the operator
$\bfD_a:=\tilD_{fv}-c_R((p(t)-a))$. Here, to simplify the
notation, we write simply $c_R(\cdot)$ for the operator
$1\otimes{c_R(\cdot)}$. Then (cf. \refl{bfD2})
\eq{bfD2}
    \bfD_a^2 \ = \ \tilD_{fv}^2 - B + |p(x)-a|^2,
\end{equation}
where $B:\Gam(\tilW,\wE)\to\Gam(\tilW,\wE)$ is a bounded operator
\footnote{The reason that, contrary to \refe{D2}, no
covariant derivatives occur in \refe{bfD2} is that we used the
right Clifford multiplication $c_R$ to define the deformed Dirac
operator $\bfD_a$. The crucial here is the fact, that $c_R$
commutes with the left Clifford multiplication $c_L$, used in the
definition of the Clifford structure on $\wE$.}.

It follows easily from \refe{bfD2} that the index $\chi_G(\bfD_a)$
is well defined and is independent of $a$, cf. \refss{indbfD}.
Moreover, $\chi_G(\bfD_a)=0$ for $a\ll 0$ and, if $a>0$ is very
large, then all the sections in $\Ker \bfD_a^2$ are concentrated
on the cylinder $M\times(0,\infty)$, not far from $M\times\{a\}$
(this part of the proof essentially repeats the arguments of
Witten in \cite{Witten82}). Hence, the calculation of
$\Ker\bfD_a^2$ is reduced to a problem on the cylinder
$M\times(0,\infty)$. It is not difficult now to show that
$\chi_G(\bfD_a)=\chi_G(D_{fv})$ for $a\gg0$, cf. \reft{a-a}.

\reft{cobordinv} follows now from the fact that $\chi_G(\bfD_a)$
is independent of $a$.

\subsection{The definition of the analytic index of a tamed
Clifford module}\Label{SS:grch} \reft{cobordinv} implies, in
particular, that, if $(\E,\v)$ is a tamed Clifford module, then
the index $\chi_G(D_{fv})$ is independent of the choice of the
admissible function $f$ and the Clifford connection on $\E$. This
proves part 2 of \reft{finite} and (cf. \refss{noncomind}) allows
us to define the {\em (analytic) index} of the tamed Clifford
module $(\E,\v)$
\[
    \chi^{\an}_G(\E,\v) \ := \ \chi_G(D_{fv}), \qquad \text{$f$ is
    an admissible function}.
\]
\reft{cobordinv} can be reformulated now as
\th{cobordinva}
The analytic indexes of cobordant tamed Clifford modules coincide.
\eth

\subsection{Index and zeros of $v$}\Label{SS:zerosv}
As a simple corollary of \reft{cobordinv}, we obtain the following
\lem{zerosv} If the vector field $v(x)\not=0$ for all $x\in M$,
then $\chi^{\an}_G(\E,\v)=0$.
\elem
\prf
Consider the product $W=M\times[0,\infty)$ and define the map
$\tv:W\to\grg$ by the formula: $\tv(x,t)=\v(x)$. Clearly,
$(W,\tv)$ is a cobordism between the tamed $G$-manifold $M$ and
the empty set. Let $\E_W$ be the lift of $\E$ to $W$. Define the
Clifford module structure $c:T^*W\to \End{\E_W}$ by the formula
\[
    c(x,a) e \ = \ c(x)e \ \pm \i a e, \qquad (x,a)\in T^*W\simeq
    T^*M\oplus\RR, \ e\in{\E_W^\pm}.
\]
Then $(\E_W,\tv)$ is a cobordism between $(\E,\v)$ and the
Clifford module over the empty set.
\eprf

\subsection{The stability of the index}\Label{SS:stDfv}
We will now amplify the above lemma and show that the index is
independent of the restriction of $(\E,\v)$ to a subset, where
$v\not=0$.

Let $(M_i,\v_i) \, i=1,2,$ be tamed $n$-dimensional $G$-manifolds.
Let $U$ be an open $n$-dimensional $G$-manifold. For each $i=1,2$,
let $\phi_i:U\to M_i$ be a smooth $G$-equivariant embedding. Set
$U_i=\phi_i(U)\subset M_i$. Assume that the boundary $\Sig_i=\d
U_i$ of $U_i$ is a smooth hypersurfaces in $M_i$. Assume also that
the vector field $v_i$ induced by $\v_i$ on $M_i$ does not vanish
anywhere on $M_i\backslash{U_i}$.

\lem{stabDv}
Let $(\E_1,\v_1)$,  $(\E_2,\v_2)$ be tamed Clifford modules over
$M_1$ and $M_2$, respectively. Suppose that the pull-backs
$\phi_i^*\E_i, \ i=1,2$ \/ are $G$-equivariantly isomorphic as
$\ZZ_2$-graded self-adjoint Clifford modules over $U$. Assume also
that $\v_1\circ\phi_1\equiv \v_2\circ\phi_2$. Then $(\E_1,\v_1)$
and $(\E_2,\v_2)$ are cobordant. In particular,
$\chi^{\an}_G(\E_1,\v_1)= \chi^{\an}_G(\E_2,\v_2)$.
\elem
The lemma is proven
\footnote{One can note that \refl{stabDv} follows immediately from
\refl{zerosv} and the additivity of the index stated in
\refc{gluing}. However, the fact that the tamed Clifford modules
$\E_1, \E_2$ of \refc{gluing} are well defined relies on
\refl{stabDv}. The lemma is also used in the proof of the
additivity formula, cf. \refs{prgluing}.}
 in \refs{prstabDv} by constructing an explicit cobordism between
$(\E_1,\v_1)$ and $(\E_2,\v_2)$.

\rem{zeros}
\refl{stabDv} implies that the index depends only on the
information near the zeros of $v$. In particular, if $G$ is a
torus and $\v:M\to\grg$ is a constant map to a generic vector of
$\grg$, this implies that the index is completely defined by the
data near the fixed points of the action. This is, essentially,
the equivariant index theorem of Atiyah-Segal-Singer (or, rather,
its extension to non-compact manifolds). See \refs{ASS} for more
details.
\erem
\vskip0.2cm

The following lemma is, in a sense, opposite to \refl{stabDv}. The
combination of these 2 lemmas might lead to an essential
simplification of a problem.
\lem{vcomp}
Let $\v_1,\v_2:M\to \grg$ be taming maps, which coincide out of a
compact subset of $M$. Then the tamed Clifford modules $(\E,\v_1)$
and $(\E,\v_2)$ are cobordant. In particular,
$\chi^{\an}_G(\E,\v_1)= \chi^{\an}_G(\E,\v_2)$.
\elem
\prf
Consider the product $W=M\times[0,1]$. Let $s:[0,1]\to[0,1]$ be a
smooth increasing function, such that $s(t)=0$ for $t\le1/3$ and
$s(t)=1$ for $t\ge2/3$. Define the map $\tv:W\to \grg$ by the
formula $\tv(x,t)=(1-s(t))\v_1(x)+s(t)\v_2(x)$. Then $(W,\tv)$ is
a cobordism between $(M,\v_1)$ and $(M,\v_2)$. Let $\E_W$ be the
lift of $\E$ to $W$, endowed with the Clifford module structure
defined in the proof of \refl{zerosv}. Then $(\E_W,\tv)$ is a
cobordism between $(\E,\v_1)$ and $(\E,\v_2)$.
\eprf

\section{The gluing formula}\Label{S:gluing}

If we cut a tamed $G$-manifold along a $G$-invariant hypersurface
$\Sig$, we obtain a manifold with boundary. By rescaling the
metric near the boundary we may convert it to a complete manifold
without boundary, in fact, to a tamed $G$-manifold. In this
section, we show that the index is invariant under this type of
surgery. In particular, if $\Sig$ divides $M$ into two pieces
$M_1$ and $M_2$, we see that the index on $M$ is equal to the sum
of the indexes on $M_1$ and $M_2$. In other words, the index is
{\em additive}. This property can be used for calculating the
index on a {\em compact manifold} $M$ (note that the manifolds
$M_1,M_2$ are non-compact even if $M$ is compact).

\subsection{The surgery}\Label{SS:surgery}
Let $(M,\v)$ be a tamed $G$-manifold. Suppose $\Sig\subset M$ is a
smooth $G$-invariant hypersurface in $M$. For simplicity, we
assume that $\Sig$ is compact. Assume also that the vector field
$v$ induced by $\v$ does not vanish anywhere on $\Sig$. Suppose
that $\E=\E^+\oplus\E^-$ is a $G$-equivariant $\ZZ_2$-graded
self-adjoint Clifford module over $M$. Denote by $\E_\Sig$ the
restriction of the $\ZZ_2$-graded Hermitian vector bundle $\E$ to
$M_\Sig:= M\backslash\Sig$.

Let $g^M$ denote the Riemannian metric on $M$. By rescaling of
$g^M$ near $\Sig$, one can obtain a complete Riemannian metric on
$M_\Sig:=M\backslash\Sig$, which makes
$(M_{\Sig},\v_{\Sig}:=\v|_{M_\Sig})$ a tamed $G$-manifold. It
follows, from the cobordism invariance of the index (more
precisely, from \refl{stabDv}), that the concrete choice of this
metric is irrelevant for our index theory. We, however, must show
that one can choose such a metric and a Clifford module structure
on $\E_\Sig$ consistently. This is done in the next subsection.

\subsection{Choice of a metric on $M_\Sig$ and a Clifford module
  structure on $\E_\Sig$}\Label{SS:chmet}
Let $\tau:M\to\RR$ be a smooth $G$-invariant function, such that
$\tau^{-1}(0)=\Sig$ and there are no critical values of $\tau$ in
the interval $[-1,1]$. Let $r:\RR\to\RR$ be a smooth function,
such that $r(t)=t^2$ for $|t|\le1/3$, $r(t)>1/9$ for $|t|>1/3$ and
$r(t)\equiv1$ for $|t|>2/3$. Set $\alp(x)=r(\tau(x))$. Define the
metric $g^{M_\Sig}$ on $M_\Sig$ by the formula
\eq{gMS}
    g^{M_\Sig} \ := \ \frac1{\alp(x)^2}\, g^M.
\end{equation}
This is a complete $G$-invariant metric on $M_\Sig$. Hence,
$(M_\Sig,g^{M_\Sig},\v_\Sig)$ is a tamed $G$-manifold.

Define a map $c_\Sig:T^*M_\Sig\to\End\E_\Sig$ by the formula
\eq{cSig}
    c_\Sig \ : = \ \alp(x) c,
\end{equation}
where $c:T^*M\to\End\E$ is the Clifford module structure on $\E$.
Then $\E_\Sig$ becomes a  $G$-equivariant $\ZZ_2$-graded
self-adjoint Clifford module over $M_\Sig$. The pair
$(\E_\Sig,\v_\Sig)$ is a tamed Clifford module.

\subsection{}\Label{SS:glu2}
Before formulating the theorem, let us make the following remark.
Suppose we choose another complete $G$-invariant metric on
$M_\Sig$ and another Clifford module structure on $\E_\Sig$, which
coincides with the ones chosen above  on $\alp^{-1}(1)\subset M$.
Then, by \refl{stabDv}, the obtained tamed Clifford module is
cobordant to  $(\E_\Sig,\v_\Sig)$. In view of this remark, we
don't demand anymore that the Clifford structure on $\E_\Sig$ is
given by \refe{cSig}.  Instead, we fix a structure of a
$G$-equivariant self-adjoint Clifford module on the bundle
$\E_\Sig$, such that ${\E_\Sig}|_{\alp^{-1}(1)}=
\E|_{\alp^{-1}(1)}$ and the corresponding Riemannian metric on
$M_\Sig$ is complete.

\th{gluing}
The tamed Clifford modules $(\E,\v)$ and $(\E_\Sig,\v_\Sig)$ are
cobordant. In particular,
\[
    \chi^{\an}_G(\E,\v) \ = \ \chi^{\an}_G(\E_\Sig,\v_\Sig).
\]
\eth
We refer to \reft{gluing} as a {\em gluing formula}, meaning that
$M$ is obtained from $M_\Sig$ by gluing along $\Sig$.

The proof of \reft{gluing} is given in \refs{prgluing}. Here we
only present the main idea of how to construct the cobordism $W$
between $M$ and $M_\Sig$.

\subsection{The idea of the proof of the gluing formula}\Label{SS:idglu}
Consider the product $M\times[0,1]$, and the set
\[
    Z \ := \ \big\{\, (x,t)\in M\times[0,1]: \, t\le 1/3, x\in \Sig\,
    \big\}.
\]
Set $W:= (M\times[0,1])\backslash{Z}$. Then $W$ is a $G$-manifold,
whose boundary is diffeomorphic to the disjoint union of
$M\backslash\Sig\simeq (M\backslash\Sig)\times\{0\}$ and $M\simeq
M\times\{1\}$. Essentially, $W$ is the required cobordism.
However, we have to be accurate in defining a complete Riemannian
metric $g^W$ on $W$, so that the condition (iii) of \refd{cobordM}
is satisfied. This is done in \refs{prgluing}.

\subsection{The additivity of the index}\Label{SS:add}
Suppose that $\Sig$ divides  $M$ into two open submanifolds $M_1$
and $M_2$, so that $M_\Sig=M_1\sqcup M_2$. The metric $g^{M_\Sig}$
induces complete $G$-invariant Riemannian metrics $g^{M_1},
g^{M_2}$ on $M_1$ and $M_2$, respectively. Let $\E_i, \v_i \
(i=1,2)$ denote the restrictions of the Clifford module $\E_\Sig$
and the taming map $\v_\Sig$ to $M_i$.  Then \reft{gluing} implies
the following
\cor{gluing}
$\displaystyle \chi^{\an}_G(\E,\v) \ = \
        \chi^{\an}_G(\E_1,\v_1) \ + \ \chi^{\an}_G(\E_2,\v_2)$.
\ecor
Thus, we see that the index of non-compact manifolds is {\em
``additive"}.

\section{The index theorem}\Label{S:IndTh}

In this section we recall the definition of the {\em topological
index} of a tamed Clifford module, cf. \cite{AtiyahTE, Paradan1},
and prove that it is equal to the analytical index.

\subsection{Transversally elliptic symbols}\Label{SS:trsym}
Let $M$ be a $G$-manifold and let $\pi:T^*M\to M$ be the
projection. A $G$-equivariant map $\sig\in
\Gam(T^*M,\Hom(\pi^*\E^+,\pi^*\E^-))$ will be called a {\em
symbol}.

Set
\[
    T^*_GM \  = \
     \big\{\, \xi\in T^*M:\, \<\xi,v(\pi(\xi))\>=0 \ \
            \text{for all} \ \ \v\in\grg\, \big\}.
\]
(Here, as usual, $v$ denotes the vector field on $M$ generated by
the infinitesimal action of $\v\in\grg$).  A symbol $\sig$ is
called {\em transversally elliptic} if
$\sig(\xi):\pi^*\E^+|_\xi\to \pi^*\E^-|_\xi$ is invertible for all
$\xi\in T^*_GM$ outside of a compact subset of \/ $T^*_GM$. A
transversally elliptic symbol defines an element of the compactly
supported $G$-equivariant $K$-theory $K_G(T^*_GM)$ of $T^*_GM$.
Thus a construction of Atiyah \cite{AtiyahTE} defines an index of
such an element. We, next, recall the main steps of this
construction.

\subsection{The index of a transversally elliptic symbol on
 a compact manifold}\Label{SS:indTEcom}
Let $\sig\in \Gam(M,\Hom(\pi^*\E^+,\pi^*\E^-))$ be a transversally
elliptic symbol on a compact manifold $M$ and let
$P:\Gam(M,\E^+)\to \Gam(M,\E^-)$ be a $G$-invariant
pseudo-differential operator, whose symbol coincides with $\sig$.

For each irreducible representation $V\in\Irr{V}$ let
\[
    \Gam(M,\E^\pm)^V \ := \
    \Hom_G(V,\Gam(M,\E^\pm))^V\otimes V
\]
be the isotipic component of $\Gam(M,\E^\pm)$ corresponding to
$V$. We denote by $P^V$ the restriction of $P$ to $\Gam(M,\E^+)^V$
so that
\[
    P^V:\, \Gam(M,\E^+)^V \ \to \ \Gam(M,\E^-)^V.
\]
It was shown by Atiyah \cite{AtiyahTE} that, if $M$ is compact,
then the operator $P^V$ is Fredholm, so that the index
\eq{indTEcom}
    \chi_G(P) \ := \
    \sum_{V\in\Irr G}\,
      \big(\, \dim\Ker P^V - \dim\Coker P^V\, \big)\cdot V
\end{equation}
is defined. Moreover, the sum \refe{indTEcom} depends only on the
(homotopy class of the) symbol $\sig$, but not on the choice of
the operator $P$. Hence, we can define the index $\chi_G(\sig)$ by
$\chi_G(\sig):=\chi_G(P)$.

\subsection{The topological index of a transversally elliptic symbol on
  a non-compact manifold}\Label{SS:indTEncom}
Let now $\sig$ be a transversally elliptic symbol on a non-compact
manifold $M$. In particular, this means that there exists an open
relatively compact subset $U\subset M$ such that $\sig(\xi)$ is
invertible for all $\xi\in \pi^{-1}(M\backslash{U})$.

Lemma~3.1 of \cite{AtiyahTE} shows that there exists a
transversally elliptic symbol $\tilsig:\pi^*\wE^+\to\pi^*\wE^-$
which represents the same element in $K_G(T^*_GM)$ as $\sig$ and
such that the restrictions of the bundles $\wE^\pm$ to
$M\backslash{U}$ are trivial, and $\tilsig|_{M\backslash{U}}$ is
an identity.

Fix an open relatively compact subset $\tilU\subset M$ which
contains the closure of $U$. Let $j:\tilU\hookrightarrow N$ be a
$G$-equivariant embedding of $\tilU$ into a compact $G$-manifold
$N$ (such an embedding always exists, cf., for example, Lemma~3.1
of \cite{Paradan1}).

The symbol $\tilsig$ extends naturally to a transversally elliptic
symbol $\tilsig_N$ over $N$. The excision theorem~3.7 of
\cite{AtiyahTE} asserts that the index $\chi_G(\tilsig_N)$ depends
only on $\sig$ but not on the choices of $U,\tilU, \tilsig$ and
$j$. One, thus, can define the {\em topological index} of $\sig$
by
\[
    \chi_G^{\top}(\sig) \ := \ \chi_G(\tilsig_N).
\]

\subsection{The topological index of a tamed Clifford module}\Label{SS:topind}
Suppose now $(\E,\v)$ is a tamed Clifford module over a complete
Riemannian manifold $M$. Clearly,
\[
    \sig_\E(\xi) \ := \ \sqrt{-1}\, c(\xi)+\sqrt{-1}\, c(v)
    \ = \ \sqrt{-1}\, c(\xi+v)
\]
defines a transversally elliptic symbol on $M$. We then define
{\em topological index of $(\E,\v)$} by
\[
    \chi_G^{\top}(\E,\v) \ := \ \chi_G^{\top}(\sig_\E).
\]

The main result of this paper is the following
\th{IndTh}
For any tamed Clifford module $(\E,\v)$ the analytic and
topological indexes coincide
\[
    \chi_G^{\an}(\E,\v) \ = \ \chi_G^{\top}(\E,\v).
\]
\eth
The proof is given in  \refs{prIndTh}. Here we only explain the
main steps of the proof.

\subsection{The sketch of the proof of \reft{IndTh}}\Label{8SS:skprIndTh}
Let $U\subset M$ be a $G$-invariant open relatively compact set
with smooth boundary which contains all the zeros of the vector
field $v$. We endow $U$ with a complete Riemannian metric and we
denote by $(\E_U,\v_U)$ the induced tamed Clifford module over
$U$.  Combining \refc{gluing} with \refl{zerosv}, we obtain
\[
    \chi^{\an}_G(\E,\v) \ = \ \chi^{\an}_G(\E_U,\v_U).
\]
Let $\tilU$ be an open relatively compact $G$-invariant subset of
$M$, which contains the closure of $U$. Fix a $G$-equivariant
embedding of $\tilU$ into a compact manifold $N$.

In \refss{topind2}, we extend $\E_U$ to a graded vector bundle
$\wE_N= \wE_N^+\oplus\wE_N^-$ over $N$ and we extend the map
$c(v)$ to a map $\oc:\wE^+\to\wE^-$, whose restriction to
$N\backslash{U}$ is the identity map.

As in \refss{chmet}, define a Clifford module $\E_U$ on $U$, which
corresponds to a complete Riemannian metric of the form
$g^U=\frac1{\alp^2}g^M$. Fix a Clifford connection $\n^{\E_U}$ on
$\E_U$ and let $f$ be an admissible function for
$(\E_U,\v|_U,\n^{\E_U})$. We can and we will assume that the
function
\[
    \of(x) \ = \
     \begin{cases}
       1/f(x), \quad x\in U;\\
       0, \quad x\not\in U,
     \end{cases}
\]
is continuous.

Let $A:\Gam(N,\tilE^+)\to \Gam(N,\tilE^+)$ denote an invertible
positive-definite self-adjoint $G$-invariant second-order
differential operator, whose symbol is equal to $|\xi|^2$. In
\refss{oper}, we show that the symbol of the transversally
elliptic operator
\[
        P \ = \ \sqrt{-1}\, \oc \ + \ \of\alp^{-1} D_U^+A^{-1/2}
\]
is homotopic to $\sig_\E$. Hence, $\chi_G^{\top}(\E,\v)=
\chi_G(P)$.

In Subsections~\ref{SS:defP} and \ref{SS:prIndTh} we use the
deformation arguments to show that $\chi_G(P)$ is equal to the
index of operator $\sqrt{-1}\oc+\of\alp^{-1}D_U^+$. Note that the
later operator is not transversally elliptic. However, an explicit
calculation made in \refss{prIndTh} shows that its index is well
defined and is equal to $\chi_G^{\an}(\E,\v)$.

\section{An example: vector bundle}\Label{S:vbundle}

In this section we assume that  \textit{$G$ is a torus} and
present a formula for the index of a tamed Clifford module over a
manifold $M$, which has a structure of the total space of a vector
bundle $p:N\to F$. This formula was probably known for a very long
time. Some particular cases can be found in
\cite[Lecture~6]{AtiyahTE} and \cite[Part~II]{Vergne96}. The
general case was proven by Paradan \cite[\S5]{Paradan1}.

The results of this section will be used in the next section to
obtain the extension of the equivariant index theorem of
Atiyah-Segal-Singer to non-compact manifolds.

\subsection{The setting}\Label{SS:setvb}
Let $M$ be the total space of a vector bundle $p:N\to F$. Assume
that the torus group  $G$ acts on $M$ by linear transformations of
the fibers and that it preserves only the zero section of the
bundle.

Let $g^M$ be a complete $G$-invariant Riemannian metric
on $M$. Let $\v:M\to\grg$ be a taming map such that both vector
fields on $M$ induced by $\v$ and by the composition
$\v\circ{p}:N\to F$ do not vanish outside of $F$.

Let $\E=\E^+\oplus\E^-$ be a $G$-equivariant $\ZZ_2$-graded
self-adjoint Clifford module over $M$.

\subsection{The decomposition of $N^\CC$}\Label{SS:decomp}
Let $N^\CC\to F$ denote the complexification of the bundle $N\to
F$. We identify $F$ with the zero section of $N$. The element
$\v(x)\in \grg \ \, (x\in F)$, acts on the fiber
$N^\CC_x:=p^{-1}(x)$ of $N^\CC$ by linear skew-adjoint
transformations. Hence, the spectrum of the restriction of the
operator $\sqrt{-1}\v(x)$ to each fiber $N^\CC$ is real.

Since $G=T^n$ does not have fixed points outside of the zero
section, the dimension of the fiber of $p:N\to F$ is even.
Moreover, we can and we will choose a $G$-invariant complex
structure $J:N\to N$ on the fibers of $N$, so that the restriction
of $\i\v(x)$ to the holomorphic space $N_x^{1,0}\subset N_x^\CC$
has only positive eigenvalues.

\subsection{The decomposition of $\E$}\Label{SS:decomp2}
Let $T_\ve{M}\subset TM$ denote the bundle of vectors tangent to
the fibers of $p:N\to F$. Let $T_\h{M}$ be the orthogonal
complement of $T_\ve{M}$. Let $T^*_\ve{M}, T^*_\h{M}$ be the dual
bundles. We have an orthogonal direct sum decomposition
$T^*M=T^*_\h{M}{\oplus}T^*_\ve{M}$. Hence, the Clifford algebra of
$T^*M$ decomposes as a tensor product
\eq{CTM}
    C(T^*M) \ = \ C(T^*_\h{M})\, {\otimes}\, C(T^*_\ve{M}).
\end{equation}

Consider the bundle $\Lam^\b((N^{0,1})^*)$ of anti-holomorphic
forms on $N$. The lift $\Lam$ of this bundle to $M$ has a natural
structure of a module over $T^*_\ve{M}$ and, in fact, is
isomorphic to the space of ``vertical spinors" on $M$, cf.
\cite[Ch.~3.2]{BeGeVe}. It follows from \cite[Ch.~3]{BeGeVe}, that
the bundle $\E$ decomposes into a (graded) tensor product
\[
    \E \ \simeq \ \calW\otimes\Lam
\]
where $\calW$ is a $G$-equivariant $\ZZ_2$-graded Hermitian vector
bundle over $M$, on which  $C(T^*_\ve{M})$ acts trivially. By
\cite[Prop.~3.27]{BeGeVe}, there is a natural isomorphism
\eq{isomvb}
    \End_{C(T^*_\ve{M})} (\calW\otimes\Lam) \ \simeq \
        \End_\CC \calW,
\end{equation}

The Clifford algebra $C(T^*_{\text{hor}}M)$ of $T^*_{\text{hor}}M$
acts on $\E$ and this action commutes with the action of
$C(T^*_\ve{M})$. The isomorphisms \refe{CTM}, \refe{isomvb} define
a $G$-equivariant action of $C(T^*_{\text{hor}}M)$ on $\calW$.

Let $S\big((N^{1,0})^*\big)=\bigoplus_k
S^k\big((N^{1,0})^*\big)\to F$ be the sum of the symmetric powers
of the dual of the bundle $N^{1,0}$. It is endowed with a natural
Hermitian metric (coming from the Riemannian metric on $M$) and
with a natural action of $G$.

\subsection{The bundle $K_F$}\Label{SS:KF}
Let us define a bundle $K_F=\W|_F\otimes S\big((N^{1,0})^*\big)$.
The group $G$ acts on $K_F$ and the subbundle of any given weight
has finite dimension. In other words,
\[
    K_F \ = \ \bigoplus_{\alp\in\calL}\, \E_\alp,
\]
where $\alp$ runs over the set of all integer weights
$\calL\simeq\ZZ^n$ of $G$ and each $\E_\alp$ is a finite
dimensional vector bundle, on which $G$ acts with weight $\alp$.
Each $\E_\alp$ is endowed with the action of the Clifford algebra
of $T^*F\simeq T^*_\h{M}|_F$, induced by its action on $\W|_F$. It
also possesses natural Hermitian metric and grading. Let $D_\alp$
denote the Dirac operator associated to a Hermitian connection on
$\E_\alp$. We will consider the (non-equivariant) index
\[
    \ind D_\alp \ = \ \dim\Ker D_\alp^+ \ - \ \dim\Ker D_\alp^-
\]
of this operator. By the Atiyah-Singer index theorem
\[
        \ind D_\alp \ = \ \int_F\, \hatA(F)\cdot \chern(\E_\alp),
\]
where $\chern(\E_\alp)$ is the Chern character of $\E_\alp$ (cf.
\cite[\S4.1]{BeGeVe}).
\th{vbundle}
  The index
  \footnote{We don't distinguish any more between the topological
  and analytic indexes in view of the index theorem~\ref{T:IndTh}.}
  of the tamed Clifford module  $(\E,\v)$ is given by
  \eq{vbundle}
       \chi_G(\E,\v) \ = \ \sum_{\alp\in\calL}\, \ind D_\alp\cdot V_\alp
       \ = \
         \sum_{\alp\in\calL}\, \left[\, \int_F\,
            \hatA(F)\cdot \chern(\E_\alp)\, \right]\cdot V_\alp,
  \end{equation}
  where $V_\alp$ denotes the (one-dimensional) irreducible
  representation of $G$ with weight $\alp$.
\eth
A $K$-theoretical proof of this theorem can be found in
\cite[\S5]{Paradan1}
\footnote{
In \cite{Paradan1}, $\v=\v\circ{p}$. The general case follows from
the fact that $(\E,\v)$ is, obviously, cobordant to
$(\E,\v\circ{p})$}. For the case when $M$ is a \ka manifold, this
theorem was proven by Wu and Zhang \cite{WuZhang} by a direct
analytic calculation of $\Ker{}D_{fv}$. The method of Wu and Zhang
works with minor changes for general manifolds. Note that our
formula is simpler than the one in \cite{WuZhang}, because we had
the freedom of choosing a convenient complex structure on $N$.
\rem{weights}
Since the action of $\i\v(x)$ on $S\big((N^{1,0})^*\big)$ has only
negative eigenvalues, there exists a constant $C>0$, such that
$\E_\alp=0$ if $\alp(\v(x))>C$ for all $x\in F\subset M$. It
follows, that $\chi_G(\E,\v)$ contains only representations with
weights $\alp$, such that $\alp(\v)\le C$.
\erem

\section{The equivariant index theorem on open manifolds}\Label{S:ASS}

In this section we present a generalization of the
Atiyah-Segal-Singer equivariant index theorem to complete
Riemannian manifolds. In particular, we obtain a new proof of the
classical Atiyah-Segal-Singer equivariant index theorem for
compact manifolds. Our proof is based on an analogue of
Guillemin-Ginzburg-Karshon linearization theorem, which, roughly
speaking, states that a tamed $G$-manifold (where $G$ is a torus)
is cobordant to the normal bundle to the fixed point set for the
$G$ action. The approach of this section is an analytic
counterpart of the $K$-theoretic study in
\cite[Lect.~6]{AtiyahTE}, \cite[Part~II]{Vergne96} and
\cite[\S4]{Paradan1}.

Throughout the section \textit{we assume that $G$ is a torus}.

\subsection{The linearization theorem}\Label{SS:linear}
Suppose $(M,\v)$ is a tamed $G$-manifold and let $F\subset M$ be
the set of points fixed by the $G$-action. Then the vector field
$v$ on $M$ vanishes on $F$. It follows that $F$ is compact. Hence,
it is a disjoint union of compact smooth manifolds $F_1\nek F_k$.
Let $N_i$ denote the normal bundle to $F_i$ in $M$ and let $N$ be
the disjoint union of $N_i$. Let $p:N\to F$ be the natural
$G$-invariant projection. In this section we do not distinguish
between the vector bundle $N$ and its total space.

Let $\v:M\to\grg$ be a taming map. Let $\v_N:N\to\grg$ be a
$G$-equivariant map, such that $\v_N|_F\equiv \v|_F$ (in
applications, we will set $\v_N=\v\circ{p}:N\to\grg$). {\em We
assume that the vector field $v$ on $M$ induced by $\v$, the
vector field $v_N$ on $N$ induced by $\v_N$ and the vector field
induced on $N$ by $\v\circ p:N\to\grg$ do not vanish outside of
$F$}. (The last condition is equivalent to the statement that $v$
has a zero of first order on $F$).

The bundles $TN|_F$ and $TM|_F$ over $F$ are naturally isomorphic.
Hence, the Riemannian metric on $M$ induces a metric on $TN|_F$.
Fix a complete $G$-invariant Riemannian metric on $N$, whose
restriction to $TN|_F$ coincides with this metric. Then $(N,\v_N)$
is a tamed $G$-manifold.

The following theorem is an analogue of Karshon's linearization
theorem \cite{Karshon98}, \cite[Ch.~4]{GGK-book}.
\begin{LinThm}\Label{T:linear}
Suppose $(M,\v)$ is  a tamed $G$-manifold, such that the vector
field induced by $\v$ on $M$ and  the vector fields induced by
$\v_N$ and $\v\circ{p}$ on $N$ do not vanish outside of $F$.
Suppose $\E, \E_N$ are  $G$-equivariant self-adjoint
$\ZZ_2$-graded Clifford modules over $M$ and $N$, respectively.
Assume that $\E_N|_F\simeq \E|_F$ as Hermitian modules over the
Clifford algebra of $T^*M|_F$. Then the tamed Clifford module
$(\E,\v)$ is cobordant to $(\E_N,\v_N)$.
\end{LinThm}
The proof is very similar to the proof of the gluing formula, cf.
\refs{prgluing}. We present only the main idea of the proof. The
interested reader can easily fill the details.

\subsection{The idea of the proof of the Linearization theorem}\Label{SS:iprlinear}
Let $V$ be a tubular neighborhood of $F$ in $M$, which is
$G$-equivariantly diffeomorphic to $N$.

Consider the product $M\times[0,1]$, and the set
\[
    Z \ := \ \big\{\, (x,t)\in M\times[0,1]: \, t\le1/3, x\not\in V\,
    \big\}.
\]
Set $W:= (M\times[0,1])\backslash{Z}$. Then $W$ is an open
$G$-manifold, whose boundary is diffeomorphic to the disjoint
union of $N\simeq V\times\{0\}$ and $M\simeq M\times\{1\}$.
Essentially, $W$ is the required cobordism. However, we have to be
accurate in defining a complete Riemannian metric $g^W$ on $W$, so
that the condition (iii) of \refd{cobordM} is satisfied. This can
be done in more or less the same way as in \refs{prgluing}.

\subsection{The equivariant index theorem}\Label{SS:ASS}
We now apply the construction of \refs{vbundle} to the normal
bundle $N_i\to F_i$. In particular, we choose a $G$-invariant
complex structure on $N$ and consider the infinite dimensional
$G$-equivariant vector bundle $K_{F_i}=\E|_{F_i}\otimes
S\big((N\ha_{i})^*\big)$. We write
$K_{F_i}=\bigoplus_{\alp\in\L}\, \E_{i,\alp}$, where $\alp$ runs
over the set of all integer weights $\L\simeq\ZZ^n$ of $G$ and
each $\E_{i,\alp}$ is a finite dimensional vector bundle on which
$G$ acts with weight $\alp$. Then, cf. \refs{vbundle}, each
$\E_{i,\alp}$ has a natural structure of a Clifford module over
$F_i$. Let $D_{i,\alp}$ denote the corresponding Dirac operator.
The main result of this section is the following analogue of the
Atiyah-Segal-Singer equivariant index theorem
\th{ASS}
Suppose the map $(M,\v)$ is  a tamed $G$-manifold, such that both
vector fields on $M$ induced by $\v$ and by $\v\circ{p}$ do not
vanish outside of $F$. Suppose $\E$ is a $\ZZ_2$-graded
self-adjoint Clifford module over $M$. Then, using the notation
introduced above, we have
\eq{ASS}
    \chi_G(\E,\v) \ = \ \sum_{\alp\in\L}\,
        \Big(\sum_{i=1}^k\, \ind D_{i,\alp}\Big)\cdot V_\alp \ = \
    \sum_{\alp\in\L}\, \Big(\sum_{i=1}^k\,
     \int_F\,
            \hatA(F_i)\cdot \chern(\E_{i,\alp})\, \Big)\cdot
            V_\alp,
\end{equation}
where $V_\alp$ denotes the (one-dimensional) irreducible
representation of $G$ with weight $\alp$.
\eth
The theorem is an immediate consequence of the cobordism
invariance of the index (\reft{cobordinv}) and the linearization
theorem~\ref{T:linear}.

\subsection{The classical Atiyah-Segal-Singer theorem}\Label{SS:ASS2}
Suppose now that $M$ is a {\em compact} manifold. Then the index
$\chi_G(D,\v)$ is independent of $\v$ and is equal to the index
representation $\chi_G(D)=\Ker D^+ \ominus \Ker D^-$. \reft{ASS}
reduces in this case to the classical Atiyah-Segal-Singer
equivariant index theorem \cite{AtSegal68}. We, thus, obtain a new
geometric proof of this theorem.


\section{An admissible function on a manifold with boundary}\Label{S:prrescaling}

In the proof of \reft{cobordinv} we will need a notion of
admissible function on a cobordism, which extends
\refd{admissible}. In this section we define this notion and prove
the existence of such a function. In particular, we will prove
\refl{rescaling} about the existence of an admissible function on
a manifold without boundary.

\subsection{}\Label{SS:admiscob}
Let $(\E,\v)$ be a tamed Clifford module over a complete
$G$-manifold $M$. Let $(W,\v_W,\phi)$ be a cobordism between
$(M,\v)$ and the empty set, cf. \refd{cobordM}. In particular, $W$
is a complete $G$-manifold with boundary and $\phi$ is a
$G$-equivariant metric preserving diffeomorphism between a
neighborhood $U$ of $\d{W}\simeq M$ and the product
$M\times[0,\eps)$.

Let $\pi:M\times[0,\eps)\to M$ be the projection. A $G$-invariant
Clifford connection $\n^{\E}$ on $\E$ induces a connection
$\n^{\pi^*\E}$ on the pull-back $\pi^*\E$, such that
\eq{tnE}
    \n^{\pi^*\E}_{(u,a)} \ := \ \pi^*\n^{\E}_u \ + \
       a\frac{\d}{\d t}, \qquad
            (u,a)\in TM\times \RR\simeq T(M\times[0,\eps)).
\end{equation}
Let $(\E_W,\v_W,\psi)$ be a cobordism between $(\E,\v)$ and the
unique Clifford module over the empty set, cf. \refd{cobordE}. In
particular, $\psi:\E_W|_U\to \pi^*\E$  is a $G$-equivariant
isometry. Let $\n^{\E_W}$ be a $G$-invariant connection on $\E_W$,
such that
$\n^{\E}|_{\phi^{-1}(M\times[0,\eps/2))}=\psi^{-1}\circ\n^{\pi^*\E}\circ\psi$.

\defe{admisb}
A smooth $G$-invariant function $f:W\to[0,\infty)$ is an
admissible function for the cobordism $(\E_W,\v_W,\n^{\E_W})$, if
it satisfies \refe{limv} and there exists a function
$h:M\to[0,\infty)$ such that $f\big(\phi^{-1}(y,t)\big)= h(y)$ for
all $y\in M, t\in[0,\eps/2)$.
\edefe

\lem{admisb}
Suppose $h$ is an admissible function for $(\E_{M},\v,\n^{\E})$.
Then there exists an admissible function $f$ on
$(\E_W,\v_W,\n^{\E_W})$ such that the restriction $f|_{M}=h$.
\elem
\prf
Consider a smooth function $r:W\to[0,\infty)$ such that
\begin{itemize}
\item $|dr(x)|\le1$, for all $x\in W$, and $\lim_{x\to\infty} r(x)=\infty$;
\item there exists a smooth function $\rho:M\to [0,\infty)$,
such that $r(\phi^{-1}(y,t))= \rho(y)$ for all $y\in M,
t\in[0,3\eps/4)$.
\end{itemize}
Then the set $\{x\in W:\, r(x)=t\}$ is compact for all $t\ge0$.
Let $v$ denote the vector field induced by $\v_W$ on $W$. Recall
that the function $\nu$ is defined in \refe{nu}. Let
$a:[0,\infty)\to[0,\infty)$ be a smooth strictly increasing
function, such that
\[
    a(t) \ \ge \  2\max\, \Big\{\,
      \frac{\nu(x)}{|v(x)|^2}: \ r(x)= t \, \Big\}+t+1; \qquad t\gg0.
\]
Let $b:[0,\infty)\to[0,\infty)$ be a smooth function, such that
\[
    0 \ < \ b(t) \ \le \ \min\, \Big\{\,
        \frac{a'(t)}{a(t)^2};\,
                 \, \frac1{t^2} \Big\}.
\]
Set 
\[
    g(t) \ = \ \Big(\, \int^\infty_t\, b(s)\, ds\,
        \Big)^{-2}.
\]
The integral converges, since $b(s)\le 1/s^2$. Moreover,
\eq{est}
    g(t)^{1/2}\ge a(t)>t; \quad g'(t)=2g^{3/2}b>0, \qquad
            t\gg0.
\end{equation}

Let $\alp:\RR\to[0,1]$ be a smooth function such that
\begin{itemize}
\item $\alp(t)=0$ for $|t|\ge 2\eps/3$;
\item $\alp(t)=1$ for $|t|\le\eps/3$.
\end{itemize}
Let $C=\max\{|\alp'(t)|:\, t\in\RR\, \}$ and let $\bet:W\to[0,1]$
be a smooth function, such that $\bet(\phi^{-1}(y,t))=\alp(t)$ for
$y\in M, t\in[0,\eps)$ and $\bet(x)=0$ for $x\not\in U$. Then
$|d\bet|\le C$.

Recall that $h$ is an admissible function for $(\E,\v,\n^{\E})$.
This function induces a function on $U\simeq{M}\times[0,\eps)$,
which, by a slight abuse of notation, we will also denote by $h$.

Set
\[
    f(x) \ := \
    \begin{cases}
            \bet(x)h(x)+(1-\bet(x))g(r(x)), \qquad &x\in
                        M\times[0,\eps),\\
            g(r(x)),    \qquad &x\not\in M\times[0,\eps).
    \end{cases}
\]
Clearly, $f\big(\phi^{-1}(y,t)\big)= h(y)$ for any $y\in M,
t\in[0,\eps/3)$.

We have to show that $\frac{f^2|v|^2}{|df||v|+f\nu+1}$ tends to
infinity as $x\to\infty$. Consider, first, the case $x\not\in U$.
Then $f(x)= g(r(x))$ and $|df|=g'|dr|\le g'$. Hence, from the
definition of the functions $a$ we get
\begin{multline}\Label{E:g<}
   \frac{|df||v|+f\nu+1}{|v|^2}
    \ \le \
    \frac{g'|v| +g\nu+1}{|v|^2}
    \\ \le \
    \big(\, g'(r)+g(r)+1\, \big)\, \frac{\nu(x)}{|v(x)|^2}
    \ \le \
    a(r)(g'(r)+g(r)+1).
\end{multline}
From \refe{g<} and \refe{est}, we obtain
\begin{multline}\Label{E:gtoinf}
    \frac{f^2|v|^2}{|df||v|+f\nu+1} \ \ge \
            \frac{g^2}{a(g'+g+1)} \  \ge \
                \frac{g^{3/2}}{2g^{3/2}b+g+1}\\
 = \ \frac{1}{2b+g^{-1/2}+g^{-3/2}} \ \ge \
    \frac{1}{2r^{-2}+r^{-1}+r^{-3}} \ \to \ \infty
\end{multline}
as $x\to\infty$. Note that \refe{gtoinf} holds even if $x\in U$,
though in this case $g(r(x))\not= f(x)$.

Consider now the case $x\in U$. Then
\begin{multline}\notag
    |df||v|+f\nu+1 \ \le \ \bet\, \big(\, |dh||v|+h\nu\, \big) \ + \
      (1-\bet)\, \big(\, g'(r)|dr||v|+g\nu\, \big) \ + \
        |d\bet|\, |h-g|\, |v| \ + \ 1\\
    \ \le \
     \bet\big(\, |dh||v|+h\nu\, \big) \ + \
       \big(\, g'|v|+g\nu\, \big) \ + \
        C(h+g)\, |v| \ + \ 1\\
    \ \le \
      2(1+C)\, \max\, \Big\{\, |dh||v|+h\nu+1; \,
            g'|v|+g\nu+1 \, \Big\}.
\end{multline}
Hence,
\begin{multline}\notag
    \frac{f^2|v|^2}{|df||v| +f\nu+1} \ \ge \
    \frac{f^2|v|^2}{2(1+C)
       \max\big\{\, |dh||v|+h\nu+1; \,
            g'|v|+g\nu+1\, \big\}} \\
    \ \ge \ \frac1{2(1+C)}\, \max\, \Big\{\,
      \frac{\bet^2 h^2|v|^2}{|dh||v|+h\nu+1}; \,
       \frac{(1-\bet)^2g^2|v|^2}{g'|v|+g\nu+1}\, \Big\}.
\end{multline}
When $x\to\infty, x\in U$, the expression $\frac{h^2|v|^2}{|dh||v|
+h\nu+1}$ tends to infinity  by definition of $h$, while
$\frac{g^2|v|^2}{g'|v| +g\nu+1}$ tends to infinity by
\refe{gtoinf}.

\refl{admisb} is proven.
\eprf

\subsection{Proof of \refl{rescaling}}\Label{SS:prrescaling}
\refl{rescaling} follows from \refl{admisb} by setting $W=M$ (so
that $\d W=\varnothing$). \hfill$\square$


\section{Proof of \reft{finite}.1}\Label{S:prfinite}


\subsection{Calculation of $D_{fv}$}\Label{SS:Du2}
Let $f$ be an admissible function and set $u=fv$. Consider the
operator
\eq{Av}
        A_u \ = \ \sum c(e_i)\, c(\nLC_{e_i}u):\, \E \ \to \ \E,
\end{equation}
where $\oe=\{e_1\ldots e_n\}$ is an orthonormal frame of $TM\simeq
T^*M$ and $\nLC$ is the Levi-Civita connection on $TM$. One easily
checks that $A_u$ is independent of the choice of $\oe$ (it
follows, also, from \refl{D2} bellow).

The proof of \reft{finite}.1 is based on the following
\lem{D2}
Let $D_u$ be the deformed Dirac operator defined in \refe{Dv},
then
\eq{D2}
    D_u^2 \ = \ D^2 \ + \ |u|^2 \ + \ {\i}A_u \ + \ {\i}\n^\E_u.
\end{equation}
\elem
The proof of the lemma is a straightforward calculation.

\subsection{Proof of \reft{finite}}\Label{SS:prfinite}
Since the operator $D_u$ is self-adjoint, $\Ker D_u=\Ker D_u^2$.
Hence, it is enough to show that each irreducible representation
of $G$ appears in $\Ker{}D_u^2$ with finite multiplicity.

Fix $V\in\Irr G$ and let
\eq{GamV}
    \Gam(M,\E)^V \  \simeq \
            \Hom_G\, \big(V,\Gam(M,\E)\big)\otimes V
\end{equation}
be the isotypic component of $\Gam(M,\E)$ corresponding to $V$.
The irreducible representation $V$ appears in $\Ker D_u^2$ with
the multiplicity equal to the dimension of the kernel of the
restriction of $D_u^2$ to the space $\Gam(M,\E)^V$. We will now
use \refe{D2} to estimate this restriction from bellow.

Note, first, that, since $\|c(v)\|=|v|$ and $\|c(e_i)\|=1$, we
have
\eq{Av2}
    \|A_u\| \ \le \ \sum_i\, \|\nLC_{e_i}u\| \ \le \
            C\, \Big(\, |df|\, |v| + f\, \|\nLC v\|\, \Big),
\end{equation}
for some constant $C>0$.

Using the definition \refe{mu} of $\mu^\E$, we obtain
$\n^\E_u=\L^\E_\u+\mu^\E(\u)$. For any $\bfa\in\grg$, the operator
$\L^\E_\bfa$ is bounded on $\Gam(M,\E)^V$. Hence, there exists a
constant $c_V$ such that
\[
    \big\|\, \L^\E_\u|_{\Gam(M,\E)^V}\, \big\| \ \le \ c_V |\u|.
\]
Thus, on $\Gam(M,\E)^V$ we have
\eq{nE<}
    \|\n^\E_u\| \ \le \ \ \|\L^\E_\u\|+\|\mu^\E(\u)\| \ = \
     f\, \big(\, \|\L^\E_\v\|+\|\mu^\E(\v)\|\, \big)
      \ \le \ f\, \big(\, c_V|\v|+\|\mu^\E(\v)\|\, \big).
\end{equation}
Combining, \refe{D2}, \refe{Av2} and \refe{nE<}, we obtain
\[
    D_u^2|_{\Gam(M,\E)^V} \ \ge \
    D^2|_{\Gam(M,\E)^V}  \ + \  f^2\, |v|^2 \ - \ \lam_V\, \Big(\,
        |df|\, |v|  \ - \ f\, \big(\, |\v|+\|\mu^\E(\v)\|+\|\nLC v\|\, \big)
        \, \Big),
\]
where $\lam_V=\max\{1,c_V,C\}$. It follows now from \refe{limv},
that there exists a real valued function $r_V(x)$ on $M$ such that
$\lim_{x\to\infty}\, r_V(x) \ = \ +\infty$ and on $\Gam(M,\E)^V$
we have
\eq{Du2a}
     D_u^2|_{\Gam(M,\E)^V} \ \ge \ D^2|_{\Gam(M,\E)^V} \ + r_V(x).
\end{equation}
It is well known (cf., for example, \cite[Lemma~6.3]{Sh7}) that
the spectrum of $D^2+r_V(x)$ is discrete. Hence,  \refe{Du2a}
implies that so is the spectrum of the restriction of $D_u^2$ to
$\Gam(M,\E)^V$. \hfill$\square$

\section{Proof of \reft{cobordinv}}\Label{S:prcobordinv}

By \refr{M12-M}, it is enough to prove \reft{cobordinv} in the
case, when $W$ is a cobordism between a tamed $G$-manifold
$(M,\v)$ and an empty set, which we shall henceforth assume.

\subsection{The Clifford module structure on $\wE$}\Label{SS:wE}
Let us consider two anti-commuting actions left and right action)
of the Clifford algebra of $\RR$ on the exterior algebra
$\Lam^\b\CC=\Lam^0\CC\oplus\Lam^1\CC$, given by the formulas
\eq{ClonR}
    c_L(t)\, \ome \ = \ t\wedge\ome \ - \ \iot_t\ome; \qquad
    c_R(t)\, \ome \ = \ t\wedge\ome \ + \ \iot_t\ome.
\end{equation}
Note, that $c_L(t)^2=-t^2$, while $c_R(t)^2=t^2$. In the
terminology of \cite{BeGeVe}, these two actions correspond to the
bilinear forms $(t,s)=ts$ and $(t,s)=-ts$ respectively.

We will use the notation of \refss{sprcobordinv}. In particular,
$\tilW$ is the manifold obtained from $W$ by attaching cylinders.
We denote by $\E'_W$ the extension of the bundle $\E_W$ to $\tilW$
and we set $\wE=\E'_W\otimes\Lam^\b\CC$.

Define a map $\tilc:T^*\tilW\to\End\wE$ by the formula
\eq{tilc}
    \tilc(v) \ := \ \i c(v)\otimes c_L(1), \qquad v\in T^*\tilW,
\end{equation}
and set
\eq{grE}
    \wE^+ \ := \ \E'_W\otimes\Lam^0; \qquad
    \wE^- \ := \ \E'_W\otimes\Lam^1.
\end{equation}
By a direct computation, one easily checks that \refe{tilc},
\refe{grE} define a structure of a  self-adjoint $\ZZ_2$-graded
Clifford module on $\wE$.

\subsection{The Dirac operator on $\tilW$}\Label{SS:tilD}
Recall that $\phi:U\to M\times[0,\eps)$ is a diffeomorphism,
defined in \refd{cobordM}, and that $\psi$ is an isomorphism
between the restriction of $\E_W$ to $U$ and the vector bundle
$\pi^*\E$ induced on $M\times[0,\eps)$ by $\E$, cf.
\refd{cobordE}. The connection $\n^\E$ on $\E$ induces a
connection $\n^{\pi^*\E}$ on $\pi^*\E$, cf. \refe{tnE}. Choose a
$G$-invariant Clifford connection on $\E_W$, whose restriction to
$\phi^{-1}(M\times[0,\eps/2))$ \/ coincides with $\n^{\pi^*\E}$.
This connection extends naturally to a $G$-invariant Clifford
connection $\n^{\wE}$ on $\wE$.

Let $\tilD$ denote the Dirac operator on $\wE$ corresponding to
the Clifford connection $\n^\tilE$. We will need an explicit
formula for the restriction of this operator to the cylinder
$M\times(0,\infty)$. Let us introduce some notation. Let
$t:M\times(0,\infty)\to(0,\infty)$ be the projection. We can and
we will view $t$ as a real valued function on the cylinder
$(0,\infty)$, so that $dt\in T^*\big(M\times(0,\infty)\big)$. Note
that $e_0:=\grad{t}\in T\big(M\times(0,\infty)\big)$ is the unit
vector tangent to the fibers of the projection
$\pi:M\times(0,\infty)\to M$. To simplify the notation, we denote
\[
    \gam \ := \ c(dt)\otimes1, \qquad
    \frac{\d}{\d t} \ = \ \n^{\wE}_{e_0}.
\]
Let $t^*D$ denote the pull-back of the operator
$D:\Gam(M,\E)\to\Gam(M,\E)$ to $M\times(0,\infty)$. Then
\eq{tilD}
    \tilD|_{M\times(0,\infty)} \ = \
    \i\, \Big(\,  t^*D+\gam\frac{\d}{\d t}\, \Big)\otimes c_L(1).
\end{equation}

\subsection{The operator $\bfD_a$}\Label{SS:bfD2}
Let $f$ be an admissible function on $M$. Fix an admissible
function on $W$ whose restriction to $M$ equals $f$, cf.
\refl{admisb}. By a slight abuse of notation, we will denote this
function by the same letter $f$. Also, to simplify the notation,
we will denote the natural extension of $f$ and $\v$ to $\tilW$ by
the same letters $f,\v$. Set
\[
    \tilD_{fv} \ = \ \tilD \ + \ {\i}\tilc(fv).
\]

Let $s:\RR\to[0,\infty)$ be a smooth function such that $s(t)=t$
for $|t|\ge1$, and $s(t)=0$ for $|t|\le1/2$. Consider the map
$p:\tilW\to\RR$ such that
\begin{align}
    p(y,t) \ &= \ s(t), \qquad &\text{for} \quad &(y,t)\in
    M\times(0,\infty);\notag\\
    p(x) \ &= \ 0, \qquad &\text{for} \quad &x\in W.\notag
\end{align}
Clearly, $p$ is a smooth function and the differential $dp$ is
uniformly bounded on $\tilW$.

By a slight abuse of notation, we will write $c_L(s)$ and $c_R(s)$
for the operators $1\otimes{}c_L(s)$ and $1\otimes{}c_R(s)$,
respectively. Note, that the operator $c_R(a)$ anti-commutes with
$\tilD_{fv}$, for any $a\in\RR$. Set
\eq{bfD}
    \bfD_a \ = \ \tilD_{fv}-c_R\big(p(x)-a\big), \qquad a\in\RR.
\end{equation}

When restricted to the cylinder, $M\times(0,\infty)$ the bundle
$\wE$ is equal to $p^*\E\otimes\Lam^\b\RR$. Let
\[
    \Pi_0:\, \wE \to p^*\E\otimes\Lam^0\RR; \qquad
    \Pi_1:\, \wE \to p^*\E\otimes\Lam^1\RR
\]
be the projections.
\lem{bfD2}
\(\displaystyle \bfD_a^2 \ = \ \tilD_{fv}^2 \ - \ B \ + \
|p(x)-a|^2,\) where $B:\wE\to\wE$ is a uniformly bounded bundle
map, whose restriction to $M\times(1,\infty)$ is equal to
$\i\gam(\Pi_1-\Pi_0)$, and whose restriction to $W$ vanishes.
\elem
\prf
Note, first, that $p(x)-a\equiv-a$ on $W$. Thus, since $c_R(a)$
anti-commutes with $\tilD_{fv}$, we have $\bfD_a^2|_W=
\tilD_{fv}^2|_W+a^2$. Hence, the identity of the lemma holds, when
restricted to $W$.

We now consider the restriction of $\bfD_a^2$ to the cylinder
$M\times(0,\infty)$. Recall that $t:M\times(0,\infty)\to
(0,\infty)$ denotes the projection and that the function $s:\RR\to
[0,\infty)$ was defined in \refss{bfD2}. Then
\[
   c_R\big(\, p(x)-a\,  \big) \ = \
    \big(\, s(t(x))-a\, \big)\,  c_R(1).
\]
Using \refe{tilD}, we obtain
\[
    \tilD_{fv}|_{M\times(0,\infty)} \ = \
       \i \Big(\, t^*D+\i c(fv)\, \Big) c_L(1)
        \ + \ \i\gam c_L(1)\frac{\d}{\d t}.
\]
The operators $\gam$ and $t^*D+\i c(fv)$ commute with
$(s(t)-a)c_R(1)$. Also the operators $c_L$ and $c_R$ anti-commute.
Hence, we obtain
\begin{multline}\notag
    \bfD_a^2 \ = \
      \tilD_{fv}^2 \ - \ \i\gam c_L(1)\frac{\d}{\d t}\, (s(t)-a)c_R(1)
        \\
          - \ \i\gam(s(t)-a)c_R(1)c_L(1)\frac{\d}{\d t} \ + \
          \Big(\, (s(t)-a)c_R(1)\, \Big)^2\\
    \ = \
        \tilD_{fv}^2 \ + \ \i s'\gam c_L(1)c_R(1) \ + \  |t-a|^2.
\end{multline}
Since $c_L(1)c_R(1)=\Pi_1-\Pi_0$, it follows, that the statement
of \refl{bfD2} holds with $B=s'\i\gam\big(\Pi_1-\Pi_0\big)$. Since
$s'=1$ on $M\times(1,\infty)$, the restriction of $B$ to this
cylinder equals $\i\gam(\Pi_1-\Pi_0)$. Finally, since $s'$ is
uniformly bounded on $\tilW$, so is the bundle map $B:\wE\to \wE$.
\eprf

Since $\bfD_a^2$ is a $G$-invariant operator, $G$ acts on
$\Ker\bfD^2_a$.
\prop{bfDfinite}
Each irreducible representation $V$ of $G$ appears in
$\Ker\bfD^2_a$ with finite multiplicity.
\eprop
\prf
We shall use the notation introduces in \refss{prfinite}. In
particular, $\Gam(\tilW,\wE)^V$ denotes the isotipic component of
$\Gam(W,\wE)$, corresponding to an irreducible representation $V$
of $G$. As in \refss{prfinite}, it is enough to prove that the
spectrum of the restriction of $\bfD_a^2$ to $\Gam(\tilW,\wE)^V$
is discrete.

The arguments of \refss{prfinite} show that there exists a smooth
function $r_V:\tilW\to[0,\infty)$ such that
\eq{tilD2}
    \tilD_{fv}^2\ge \tilD^2+r_V(x)
\end{equation}
on $\Gam(\tilW,\wE)^V$ and the following 2 conditions hold
\begin{itemize}
\item $r_V(x)\to+\infty$ as $x\to\infty$ and  $x\in W$;
\item $r_V\big(\, y,t\, \big)\to+\infty$ uniformly in $t\in[0,\eps)$, as $y\in
M$ and $y\to\infty$.
\end{itemize}

Let $\|B(x)\|, \, x\in\tilW$ denote the norm of the bundle map
$B_x:\wE_x\to \wE_x$ and let $\|B\|_\infty=
\sup_{x\in\tilW}\|B(x\|$. Set
\eq{R}
    R_V(x) \ := \ r_V(x) \ + \ |p(x)-a|^2 \ - \ \|B\|_\infty.
\end{equation}
Then $R_V(x)\to+\infty$ as $\tilW\ni{x}\to\infty$. Also, by
\refl{bfD2} and \refe{tilD2}, we have
\eq{bfD2>}
    \bfD_a^2|_{\Gam(\tilW,\wE)^V} \ \ge \
            \tilD^2 \ + \ R_V(x).
\end{equation}
By  \cite[Lemma~6.3]{Sh7}), the spectrum of $\tilD^2+R_V(x)$ is
discrete. Hence, \refe{bfD2>} implies that so is the spectrum  of
the restriction of $\bfD_a^2$ to $\Gam(\tilW,\wE)^V$.
\eprf

\subsection{The index of $\bfD_a$}\Label{SS:indbfD}
If $V$ is an irreducible representation of $G$, we denote by
$\bfD_a^{V,\pm}$ the restriction of $\bfD_a$ to the space
$\Gam(\tilW,\wE^\pm)^V$. It follows from \refp{bfDfinite}, that
$\bfD_a^{V,\pm}$ is a Fredholm operator. In particular, all the
irreducible representations of $G$ appear in $\Ker\bfD_a$ with
finite multiplicities.  Hence, we can define the index
$\chi_G(\bfD_a)$ using \refe{char}, or, equivalently, by the
formula
\eq{charbfD}
    \chi_G(\bfD_a) \ := \
        \sum_{V\in\Irr G}\, \Big(\,
            \dim\Ker\bfD_a^{V,+} \ - \
            \dim\Ker\bfD_a^{V,-}\, \Big)\, V.
\end{equation}

\prop{depofa}
\(\displaystyle\chi_G(\bfD_a)=0 \) for all $a\in\RR$.
\eprop
\prf
Each summand in \refe{charbfD} is the index of the operator
$\bfD_a^{V,+}$. Thus, since
\[
    \bfD_a^{V,+}-\bfD_b^{V,+}=c_R(b-a):\,
        L_2(\tilW,\wE) \ \to \ L_2(\tilW,\wE)
\]
is bounded operator depending continuously on $a,b\in\RR$, the
index $\chi_G(\bfD_a)$ is independent of $a$.

Therefore, it is enough to prove the proposition for one
particular value of $a$. Recall that the norm $\|B\|_\infty$ was
defined in the proof of \refp{bfDfinite}. Choose $a\ll0$ such that
$a^2>\|B\|_\infty$. It follows now from \refe{R} and \refe{bfD2>},
that $\bfD_a^2>0$, so that $\Ker\bfD^2_a=0$. Hence,
$\chi_G(\bfD^2_a)=0$.
\eprf

\reft{cobordinv} follows now from \refp{depofa} and the following
\th{a-a}
\(\displaystyle \chi_G(\bfD_a) \ = \
         \chi_G(D_{fv}) \) for $a\gg0$.
\eth
The proof of the theorem occupies the next section.


\section{Proof of \reft{a-a}}\Label{S:pra-a}

\subsection{The plan of the proof}\Label{SS:plana-a}
We consider an operator $\mD$ on the cylinder $M\times\RR$, with
the following property: Let $\mD_a, \, a\in\RR$ denote the
operator obtained from $\mD$  by the shift $T_a:(x,t)\to (x,t+a)$
(see \refss{DaV} for a precise definition). Then the restrictions
of $\mD_a$ and $\bfD_a$ to the cylinder $M\times(1,\infty)$
coincide. Following Shubin \cite{Sh5}, we call $\mD$ the {\em
model} operator.

In \refl{model}, we show that $\chi_G(D_{fv})=\chi_G(\mD_a)$ for
any $a\in\RR$.

The explicit formula for $\bfD_a^2$, obtained in \refl{bfD2},
shows that the restriction of this operator to the compliment of
$M\times(1,\infty)$ becomes ``very large" as $a\to\infty$. It
follows that the eigenfunctions of $\bfD_a$ are concentrated on
$M\times(1,\infty)$ for large $a$. Hence,  the kernel of $\bfD_a$
can be estimated using the calculations on this cylinder, i.e., in
terms of $\mD_a$. This is done in \refp{N}. Using this proposition
it is easy to show that $\chi_G(\bfD_a)=\chi_G(\mD_a)$ for large
$a$, cf. \refss{pra-a}.

\subsection{The model operator on the cylinder}\Label{SS:model}
The restriction of $\wE$ to $M\times(0,\infty)$ extends naturally
to a Hermitian vector bundle over $M\times\RR$, which we will also
denote by $\wE$. If $t:M\times\RR\to \RR$ denotes the projection,
then $\wE\simeq t^*\E\otimes\Lam^\b\CC$. Define the Clifford
module structure and the grading on $\wE$ using \refe{tilc},
\refe{grE}.

Let $\tilD, t^*D:\Gam(M\times\RR,\wE)\to \Gam(M\times\RR,\wE)$ be,
correspondingly, the Dirac operator on $\wE$ and the pull-back of
$D$. Using the notation introduced in  \refss{tilD}, we can write
\[
    \tilD \ = \
    \i\, \Big(\,  t^*D+\gam\frac{\d}{\d t}\, \Big)\otimes c_L(1).
\]
Set
\eq{model}
    \mD \ : = \ \tilD \ + \ \i\tilc(fv) \ - \ c_R\big(\, t(x)\, \big):\,
        \Gam(M\times\RR,\wE) \ \to \ \Gam(M\times\RR,\wE).
\end{equation}
We will refer to $\mD$ as the {\em model operator}, cf.
\cite{Sh5}. This is a $G$-invariant elliptic operator. Moreover,
it follows from \refp{bfDfinite} that the index of $\mD$ is well
defined. To see this, one can set $W=M\times[0,1]$ in
\refp{bfDfinite}, and view $M\times\RR$ as a manifold obtained
from $W$ by attaching cylinders.
\lem{model}
The kernel of the model operator $\mD$ is $G$-equivariantly
isomorphic (as a graded space) to $\Ker(D_{fv})$. In particular,
the index $\chi_G(\mD)$ is well defined and is equal to
$\chi_G(D_{fv})$.
\elem
\prf
The same calculations as in the proof of \refl{bfD2}, show that
\[
    (\mD)^2 \ = \ t^*\Big(\, D + c(fv)\, \Big)^2 \ + \
    \Big(\, -\frac{\d^2}{\d t^2}- \i\gam(\Pi_1-\Pi_0)+t^2
    \, \Big).
\]
Thus, we obtain the following formulas for the restrictions of
$(\mD)^2$ to the spaces
$\Gam(M\times\RR,\E^\pm\otimes\Lam^\b\CC)$:
\eq{Dmodpm}
    (\mD)^2|_{\Gam(M\times\RR,\E^\pm\otimes\Lam^\b\CC)} \ = \
    t^*\Big(\, D + c(fv)\, \Big)^2 \ + \
    \Big(\, -\frac{\d^2}{\d t^2}\pm(\Pi_1-\Pi_0)+t^2
    \, \Big).
\end{equation}
Here the first summand coincides with the lift of $D_{fv}^2$ to
$\wE$, while the second summand may be considered as an operator
acting on the space of $\Lam^\b\CC$-valued functions on $\RR$.
Also, both summands in the right hand side of \refe{Dmodpm} are
non-negative. Hence, the kernel of $(\mD)^2$ equals the tensor
product of the kernels of these two operators.

The space $\Ker\big(-\frac{\d^2}{\d t^2}+\Pi_1-\Pi_0+t^2\big)$ is
one dimensional and is spanned by the function
$\alp(t):=e^{-t^2/2}\in \Lam^0\RR$. Similarly,
$\Ker\big(-\frac{\d^2}{\d t^2}+\Pi_0-\Pi_1+t^2\big)$ is one
dimensional and is spanned by the one-form
$\bet(t):=e^{-t^2/2}ds$, where we denote by $ds$ the generator of
$\Lam^1\CC$. It follows that
\begin{align}
    \Ker(\mD)^2|_{\Gam(M\times\RR,\E^+\otimes\Lam^\b\CC))} \ &\simeq \
    \Big\{\, t^*\sig\otimes\alp(t): \
    \sig\in\Ker D_{fv}^2|_{\Gam(M,\E^+)}\, \Big\}; \notag\\
    \Ker(\mD)^2|_{\Gam(M\times\RR,\E^-\otimes\Lam^\b\CC)} \ &\simeq \
    \Big\{\, t^*\sig\otimes\bet(t): \
    \sig\in\Ker D_{fv}^2|_{\Gam(M,\E^-)}\, \Big\}. \notag
\end{align}
\eprf

\subsection{}\Label{SS:DaV}
Let $T_a:M\times\RR\to M\times\RR, \ T_a(x,t)=(x,t+a)$ be the
translation. Using the trivialization of $\wE$ along the fibers of
$t:M\times\RR\to\RR$, we define the pull-back map
$T_a^*:\Gam(M\times\RR,\wE)\to \Gam(M\times\RR,\wE)$. Set
\[
    \mD_a \ := \ T_{-a}^*\circ\mD\circ T^*_a \ = \
        \tilD \ + \ \i \tilc(fv) \ - \ c_R\big(\, t(x)-a\, \big)
\]
Then $\chi_G(\mD_a)=\chi_G(\mD)$, for any $a\in\RR$.

\subsection{}\Label{SS:NlamA}
If $A$ is a self-adjoint operator with discrete spectrum and
$\lam\in\RR$, we denote by $N(\lam,A)$ the number of the
eigenvalues of $A$ not exceeding $\lam$ (counting multiplicities).

Recall from \refss{indbfD}, that $\bfD_a^{V,\pm}$ denote the
restriction of $\bfD_a$ to the space $\Gam(\tilW,\tilE^\pm)^V$.
Similarly, let $\mD_{V,\pm}, \mD_{V,\pm,a}$ denote the restriction
of the operators $\mD, \mD_a$  to the spaces
$\Gam(M\times\RR,\wE^\pm)^V$.
\prop{N}
Let $\lam_{V,\pm}$ denote the smallest non-zero eigenvalue of
$(\mD_{V,\pm})^2$. Then, for any $\eps>0$, there exists
$A=A(\eps,V)>0$, such that
\eq{N}
    N\big(\lam_{V,\pm}-\eps, (\bfD_a^{V,\pm})^2\big)
    \ = \ \dim\Ker(\mD_{V,\pm})^2,
\end{equation}
for any $a>A$.
\eprop
Before proving the proposition let us explain how it implies
\reft{a-a}.

\subsection{Proof of \reft{a-a}}\Label{SS:pra-a}
Let $V$ be an irreducible representation of $G$ and let $\eps$ and
$a$ be as in \refp{N}. Let $E^{V,\pm}_{\eps,a}\subset
\Gam(\tilW,\wE^\pm)^V$ denote the vector space spanned by the
eigenvectors of the operator $(\bfD^{V,\pm})^2_a$ with eigenvalues
smaller or equal to $\lam_{V,\pm}-\eps$. The operator
$\bfD^{V,\pm}_a$ sends $E^{V,\pm}_{\eps,a}$ into
$E^{V,\mp}_{\eps,a}$. Since the dimension of the space
$E^{V,\pm}_{\eps,a}$ is finite, it follows that
\[
    \dim\Ker\bfD_a^{V,+} \ - \
            \dim\Ker\bfD_a^{V,-}  \ = \
    \dim E^{V,+}_{\eps,a} \ - \
            \dim E^{V,-}_{\eps,a}.
\]
By \refp{N}, the right hand side of this equality equals
$\dim\Ker\mD_{V,+}-\dim\Ker\mD_{V,+}$. Thus
\[
    \chi_G(\bfD_a) \ = \  \chi_G(\mD).
\]
\reft{a-a} follows now from \refl{model}. \hfil$\square$

The rest of this section is occupied with the proof of \refp{N}.

\subsection{Estimate from above on $N(\lam_{V,\pm}-\eps,(\bfD_a^{V,\pm})^2)$}\Label{SS:above}
We will first show that
\eq{above}
       N(\lam_{V,\pm}-\eps,(\bfD_a^{V,\pm)^2}))\le\dim\Ker\mD_{V,\pm}.
\end{equation}
To this end we will estimate the operator $\H$ from below. We will
use the technique of   \cite{sh2,BrFar1}, adding some necessary
modifications.

\subsection{The IMS localization}\Label{SS:IMS}
Let $j,\oj:\RR\to[0,1]$ be  smooth functions such that
$j^2+\oj^2\equiv0$ and $j(t)=1$ for $t\ge3$, while $j(t)=0$ for
$t\le 2$.

Recall that $t:M\times\RR\to\RR$ denote the projection and that
the map $p:\tilW\to\RR$ was defined in \refss{bfD2}. For each
$a>0$, define smooth functions $J_a$ and $\oJ_a$ on $M\times\RR$
by the formulas:
\[
    J_a(x)=j(a^{-1/2}t(x)),\qquad \oJ_a(x)=\oj(a^{-1/2}t(x)).
\]
By a slight abuse of notation we will denote by the same letters
the smooth functions on $\tilW$ given by the formulas
\[
    J_a(x)=j(a^{-1/2}p(x)),\qquad \oJ_a(x)=\oj(a^{-1/2}p(x)).
\]

We identify the functions $J_a,\oJ_a$ with the corresponding
multiplication operators. For  operators  $A,B$, we denote by
$[A,B]=AB-BA$ their commutator.

The following version of IMS localization formula (cf.
\cite{cfks}) is due to Shubin \cite[Lemma 3.1]{Sh5} (see also
\cite[Lemma~4.10]{BrFar1}).
\lem{sh1}The following operator identity holds
  \eq{sh1}
        \H=\oJ_a \H\oJ_a+J_a\H J_a+\frac12[\oJ_a,[\oJ_a,\H]]+\frac12[J_a,[J_a,\H]].
  \end{equation}
\elem
\prf
  Using the equality $J_a^2+\oJ_a^2=1$ we can write
  $$
        \H=J_a^2\H+\oJ_a^2\H=J_a\H J_a +\oJ_a\H\oJ_a+J_a[J_a,\H]+\oJ_a[\oJ_a,\H].
  $$
   Similarly,
  $$
        \H=\H J_a^2+\H\oJ_a^2=J_a\H J_a +\oJ_a\H\oJ_a-[J_a,\H]J_a-[\oJ_a,\H]\oJ_a.
  $$
  Summing these identities and dividing by 2, we come to \refe{sh1}.
\eprf
We will now estimate each of the summands in the right hand side
of \refe{sh1}.

\lem{sh3}
  There exists $A>0$, such that
  \eq{sh3}
        \oJ_a\H\oJ_a\ge \frac{a^2}8\oJ_a^2,
  \end{equation}
  for any $a>A$.
\elem
\prf
   Note that $p(x)\le3 a^{1/2}$ for any $x$ in the support of
   $\oJ_a$. Hence, if $a>36$, we have $\oJ_a^2|p(x)-a|^2\ge
   \frac{a^2}4\oJ_a^2$.

   Recall that the norm $\|B\|_\infty$  was defined in the proof of
   \refp{bfDfinite}. Set
   \[
      A \ = \  \max\big\{\, 36, 4\|B\|_\infty^{1/2}\, \big\}
   \]
   and let $a>A$. Using \refl{bfD2}, we obtain
   \[
        \oJ_a\H\oJ_a \ \ge \oJ_a^2 |p(x)-a|^2 \ - \ \oJ_a B\oJ_a
        \ \ge \ \frac{a^2}4\oJ_a^2 \ - \ \oJ_a^2\|B\|_\infty
        \ \ge \ \frac{a^2}8\oJ_a^2.
   \]
\eprf
\subsection{}\Label{SS:Pa}
Let $P_{a}:L_2(M\times\RR,\wE)\to \Ker\mD_a$ be the orthogonal
projection. Let $P^{V,\pm}_a$ denote the restriction of $P_a$ to
the space $L_2(M\times\RR,\wE^\pm)^V$. Then $P^{V,\pm}_a$ is a
finite rank operator and its rank equals $\dim\Ker\mD_{V,\pm,a}$.
Clearly,
\eq{ge}
        \mD_{V,\pm,a}+\lam_{V,\pm}P^{V,\pm}_a\ge \lam_{V,\pm}.
\end{equation}
By identifying the support of $J_a$ in $M\times\RR$ with a subset
of $\tilW$, we can and we will consider $J_aP_aJ_a$ and
$J_a\mD_aJ_a$ as operators on $\tilW$. Then $J_a\H
J_a=J_a\mD_aJ_a$. Hence, \refe{ge} implies the following
\lem{local}
  For any $a>0$,
  \eq{local}
        J_a\bfD_a^{V,\pm} J_a+
        \lam_{V,\pm}J_aP^{V,\pm}_aJ_a\ge \lam_{V,\pm}J_a^2, \qquad
                \rk J_aP^{V,\pm}_aJ_a\le \dim\Ker\mD_{V,\pm}.
  \end{equation}
\elem
For an operator $A:L_2(\tilW,\wE)\to L_2(\tilW,\wE)$, we denote by
$\|A\|$ its norm with respect to $L_2$ scalar product on
$L_2(\tilW,\wE)$.
\lem{sh2}Let \/ $\displaystyle
   C \ = \ 2\max\, \Big\{\, \max\{|dj(t)|^2,|d\oj(t)|^2\}: \,
        t\in\RR\, \Big\}$.
  Then
  \eq{sh2}
        \|[J_a,[J_a,\H]]\|\le Ca^{-1}, \qquad   \|[\oJ_a,[\oJ_a,\H]]\|\le
        Ca^{-1}, \qquad \text{for any} \quad a>0.
  \end{equation}
\elem
\prf
Since $\H$ is a Dirac operator, it follows from
\cite[Prop.~2.3]{BeGeVe}, that
\[
    [J_a,[J_a,\H]] \ = \ -2|dJ_a|^2,
    \qquad [\oJ_a,[\oJ_a,\H]] \ = \ -2|d\oJ_a|^2.
\]
The lemma follows now from the obvious identities
\[
    |dJ_a(x)| \ = \
       a^{-1/2}\big|\, dj(a^{-1/2}p(x))\, \big|, \qquad
    |d\oJ_a(x)| \ = \ a^{-1/2}\big|\, d\oj(a^{-1/2}p(x))\, \big|.
\]
\eprf

From Lemmas~\ref{L:sh1}, \ref{L:local} and \ref{L:sh2} we obtain
the following
\cor{Da>}
For any $\eps>0$, there exists $A=A(\eps,V)>0$, such that, for any
$a>A$, we have
\eq{Da>}
    \bfD_a^{V,\pm}+\lam_{V,\pm}J_aP_a^{V,\pm}J_a \ \ge \ \lam_{V,\pm}-\eps,
    \qquad
     \rk J_aP^{V,\pm}_aJ_a\le \dim\Ker\mD_{V,\pm}.
\end{equation}
\ecor

The estimate \refe{above} follows now from \refc{Da>} and the
following general lemma \cite[p. 270]{ReSi4}.
\lem{general}
   Assume that $A, B$ are self-adjoint operators in a Hilbert space
   $\calH$ such that  $\rk B\le k$ and there exists $\mu>0$
   such that
   $$
        \langle (A+B)u,u\rangle \ge \mu\langle u,u\rangle
                \quad \text{for any} \quad u\in\Dom(A).
   $$
   Then $N(\mu -\eps, A)\le k$ for any $\eps>0$.
\elem

\subsection{Estimate from below on $N(\lam_{V,\pm}-\eps,(\bfD_a^{V,\pm})^2)$}\Label{SS:below}
To prove \refp{N} it remains now to show that
\eq{below}
       N(\lam_{V,\pm}-\eps,(\bfD_a^{V,\pm})^2) \ \ge \ \dim\Ker\mD_{V,\pm}.
\end{equation}
Let $E^{V,\pm}_{\eps,a}\subset \Gam(\tilW,\wE)$ denote the vector
space spanned by the eigenvectors of the operator
$(\bfD^{V,\pm}_a)^2$ with eigenvalues smaller or equal to
$\lam_{V,\pm}-\eps$. Let
$\Pi^{V,\pm}_{\eps,a}:L_2(\tilW,\wE^\pm)^V\to E^{V,\pm}_{\eps,a}$
be the orthogonal projection.  Then
\eq{rkPi}
        \rk \Pi^{V,\pm}_{\eps,a} \ = \ N(\lam_{V,\pm}-\eps,(\bfD_a^{V,\pm})^2).
\end{equation}
As in \refss{Pa}, we can and we will consider
$J_a\Pi^{V,\pm}_{\eps,a}J_a$ as an operator on
$L_2(M\times\RR,\wE^\pm)^V$. The proof of the following lemma does
not differ from the proof of \refc{Da>}.
\lem{below}
For any $\eps>0$, there exists $A=A(\eps,V)>0$, such that, for any
$a>A$, we have
\eq{Da<}
    \mD_{V,\pm,a}+\lam_{V,\pm}J_a\Pi_a^{V,\pm}J_a \ \ge \ \lam_{V,\pm}-\eps,
    \qquad
     \rk J_a\Pi^{V,\pm}_aJ_a\le \dim N(\lam_{V,\pm}-\eps,(\bfD_a^{V,\pm})^2).
\end{equation}
\elem
The estimate \refe{below} follows now from \refe{rkPi},
\refl{below} and \refl{general}.

The proof of \refp{N} is complete. \hfil$\square$

\section{Proof of \refl{stabDv}}\Label{S:prstabDv}

\subsection{The restriction of the Clifford module to $U$}\Label{SS:restU}
Recall that $U_1=\phi_1(U)$. To simplify the notation we identify
$U_1$ with $U$ and write $U=U_1$. We also denote the boundary
$\d{U}$ of $U$ in $M_1$ by $\Sig$. Recall that it is a smooth
$G$-invariant hypersurface in $M_1$.

Let $\E_{U}, \v_{U}$ denote the restrictions of $\E_1$ and $\v_1$
to $U$, respectively. We will define a structure of a tamed
Clifford module on $\E_U$. For this we need to change the Clifford
action of $T^*U$ on $\E_U$, so that the corresponding Riemannian
metric on $U$ is complete.

Let $\alp:M_1\to\RR$ be a smooth $G$-invariant function, such that
$0$ is a regular value of $\alp$ and $\alp^{-1}((0,\infty))=U, \,
\alp^{-1}(0)=\Sig$.

Let $c_1:T^*M_1\to\End\E_1$ denote the Clifford module structure
on $\E_1$. Define a map $c_U:T^*U\to\End\E_U$ by the formula
\[
    c_U(a) \ := \alp(x) c_1(a), \qquad a\in T^*_xU.
\]
Then $c_U$ defines a Clifford module structure on $\E_U$, which
corresponds to the Riemannian metric $g^U=\alp^{-2}{g^M|_U}$,
which is complete. From now on we denote by $\E_U$ the Clifford
module defined by $c_U$. We also endow $\E_U$ with the Hermitian
structure obtained by the restriction of the Hermitian structure
on $\E_1$. Then $(\E_U,\v_U)$ is a tamed Clifford module. Clearly,
to prove \refl{stabDv}, it is enough to show that this module is
cobordant to $(\E_1,\v_1)$.

\subsection{Proof of \refl{stabDv}}\Label{SS:prstabDv}
Since we will not work with $M_2,\E_2$ any more, we will simplify
the notation by omitting the subscript ``1" everywhere. Thus we
set $M=M_1, \E=\E_1$, etc.  We will construct now a cobordism
between $\E_U$ and $\E$.

Consider the product $M\times[0,1]$, and the set
\[
    Z \ := \ \big\{\, (x,t)\in M\times[0,1]: \, t\le1/3, x\not\in U\,
    \big\}.
\]
Set $W:= (M\times[0,1])\backslash{Z}$. Then $W$ is a non-compact
$G$-manifold, whose boundary is diffeomorphic to the disjoint
union of $U\simeq U\times\{0\}$ and $M\simeq M\times\{1\}$.
Essentially, $W$ is the required cobordism, but we need to define
all the structures on $W$.

Let $\mu,\nu:W\to(0,\infty)$ be a smooth $G$-invariant functions
such that
\begin{itemize}
\item $\mu(x,t)=1$, if $t\ge2/3$;
\item $\mu(x,t)= 1/\alp(x)$, if $t\le 1/2$ and $\alp(x)\ge t-1/3$;
\end{itemize}
and
\begin{itemize}
\item $\nu(x,t)=1$, if $t\ge2/3$ or $t\le1/4$;
\item $\nu(x,t)=1/t$ if $t\le 1/2$ and $t-1/3\ge\alp(x)$.
\end{itemize}
Define the metric $g^W$ on $W$ by the formula
\[
    g^W\Big(\, (\xi_1,\tau_1),\, (\xi_2,\tau_2) \, \Big)
    \ := \
    \mu(x,t)^{2}\, g^{M}(\xi_1,\xi_2)
    \ + \ \nu(x,t)^2\tau_1\tau_2,
\]
where $(\xi_1,\tau_1), (\xi_2,\tau_2)\in T_xM\oplus\RR\simeq
T_{(x,t)}W$.  Then $g^W$ is a complete $G$-invariant metric.

Consider the $G$-invariant neighborhood
\eq{calO}
    \calO \ := \
    \Big\{\, (x,t):\, \,  4t < \alp(x)    \, \Big\}\, \bigsqcup \,
      \Big\{\, (x,t):\, x\in M, 3/4< t\le 1\, \Big\}
\end{equation}
of $\d W$. Define a map
$\phi:\big(U\times[0,1/4)\big)\sqcup\big(M\times(-1/4,0]\big)\to
\calO$ by the formulas
\begin{alignat}{2}
    \phi(x,t) \ &:= \ (x, t),
    \qquad\qquad &x\in U, \ 0\le t<1/4; \notag\\
    \phi(x,t) \ &:= \ (x,1+t), \qquad\qquad &x\in M, \ -1/4< t\le 0.
    \Label{E:phi}
\end{alignat}
Clearly, $\phi$ is a $G$-equivariant metric preserving
diffeomorphism, satisfying condition (iii) of \refd{cobordM}.
Define a map $\v_W:W\to\grg$ by the formula $\v_W(x,t)=\v(x)$.
Then $(W,\v_W,\phi)$ is a cobordism between $(M,\v)$ and
$(U,\v_U)$.

Let $\E_W$ be the $G$-equivariant Hermitian vector bundle on $W$,
obtained by restricting to $W$ of the pull-back of $\E$ to
$M\times[0,1]$. Define the map $c_W:T^*W\to\End\E_W$ by the
formula
\[
    c_W(a,b) e
    \ = \
    \mu(x,t)^{-1} c(a) e
    \ \pm \ \nu(x,t)^{-1}b\i e,
    \qquad
      e\in\E_{W,(x,t)}^\pm, \
      (a,b)\in T^*_xM\oplus\RR\simeq T^*_{(x,t)}W.
\]
Then $c_W$ defines a structure of a $G$-equivariant self-adjoint
Clifford module on $\E_W$, compatible with the Riemannian metric
$g^W$, whose restriction to $U\times\{0\}\subset W$ is isomorphic
to $\E_U$ and whose restriction to $M\times\{1\}$ is isomorphic to
$\E$.

One easily checks that the tamed Clifford module $(\E_W,\v_W)$
provides a cobordism between $(\E,\v)$ and $(\E_U,\v_U)$.
\hfill$\square$

\section{Proof of the gluing formula}\Label{S:prgluing}

\subsection{A cobordism between $M$ and $M_\Sig$}\Label{SS:M-MS}
Consider the product $M\times[0,1]$, and the set
\[
    Z \ := \ \big\{\, (x,t)\in M\times[0,1]: \, t\le 1/3, x\in \Sig\,
    \big\}.
\]
Set $W:= (M\times[0,1])\backslash{Z}$. Then $W$ is an open
$G$-manifold, whose boundary is diffeomorphic to the disjoint
union of $M\backslash\Sig\simeq (M\backslash\Sig)\times\{0\}$ and
$M\simeq M\times\{1\}$. Essentially, $W$ is the required
cobordism. However, we have to be accurate in defining a complete
Riemannian metric $g^W$ on $W$, so that the condition (iii) of
\refd{cobordM} is satisfied.

Recall that the function $\alp:M\to[0,1]$ was defined in
\refss{surgery}. Let the function $s:W\to(0,\infty)$ and the
metric $g^W$ on $W$ be as in \refss{prstabDv}. The group $G$ acts
naturally on $W$ preserving the metric $g^W$. This makes $W$ a
complete $G$-manifold with boundary. Define a $G$-equivariant map
$\v_W:W\to\grg$ by the formula $\v_W(x,t)=\v(x)$.

We still have some freedom of choosing a Riemannian metric on
$M_\Sig$ and a Clifford module structure on $\E_\Sig$, cf.
\refl{stabDv}. To make these choices, consider a map
$\varphi:M_\Sig\to\d{W}$ defined by
\[
    \varphi(x) \ = \ \big(\, x,0\, \big).
\]
Let $g^{M_\Sig}$ be the pull-back to $M_\Sig$ of the metric $g^W$.
Then $(M_\Sig,g^{M_\Sig},\v_\Sig)$ is a tamed $G$-manifold.

Let $\calO$ be a $G$-invariant neighborhood of $\d W$, defined by
\refe{calO}. Define a map
$\phi:\big(M_\Sig\times[0,1/4)\big)\sqcup\big(M\times(-1/4,0]\big)\to
\calO$ by \refe{phi}.
Then $\phi$ is a $G$-equivariant metric preserving diffeomorphism,
satisfying condition (iii) of \refd{cobordM}. One easily checks
that $(W,\v_W,\phi)$ is a cobordism between $(M,\v)$ and
$(M_\Sig,\v_\Sig)$.

\subsection{The bundle $\E_W$. Proof of \reft{gluing}}\Label{SS:prgluing}
Consider the Clifford module $\E_W$ over $W$ defined as in
\refss{prstabDv}. Then the restriction of $\E_W$ to
$M\times\{0\}\subset W$ is isomorphic to $\E$.

Recall that $\varphi:M_\Sig\to W$ is a diffeomorphism of $M_\Sig$
onto a piece of boundary of $W$. Set $\E_{M_\Sig}=\varphi^*\E_W$.
Clearly, $\E_{M_\Sig}$ is a $G$-equivariant $\ZZ_2$-graded
self-adjoint Clifford module over the Riemannian manifold
$(M_\Sig,g^{M_\Sig})$. Moreover, the restriction of $\E_\Sig$ to
$\alp^{-1}(1)$ equals $\E|_{\alp^{-1}(1)}$.

The tamed Clifford module $(\E_W,\v_W)$ provides a cobordism
between $(\E,\v)$ and $(\E_\Sig,\v_\Sig)$. \hfill$\square$

\section{Proof of the index theorem}\Label{S:prIndTh}

\subsection{A tamed Clifford module over $U$}\Label{SS:EU}
First, we define a complete metric on $U$ and a tamed Clifford
module over $U$, using the construction of \refss{chmet}.

Let $\tau:M\to\RR$ be a smooth $G$-invariant function such that
$\tau^{-1}((0,\infty))=U$, $\tau^{-1}(0)=\partial{U}$ and there
are no critical values of $\tau$ in the interval $[-1,1]$.  Let
$r:\RR\to\RR$ be a smooth function, such that $r(t)=t^2$ for
$|t|\le1/3$, $r(t)>1/9$ for $|t|>1/3$ and $r(t)\equiv1$ for
$|t|>2/3$. Set $\alp(x)=r(\tau(x))$. Define a complete
$G$-invariant metric $g^{U}$ on $U$ by the formula
\[
    g^{U} \ := \ \frac1{\alp(x)^2}\, g^M|_U.
\]
Define a map $c_U:T^*U\to\End\E|_U$ by the formula
\[
    c_U \ : = \ \alp(x) c,
\]
where $c:T^*M\to\End\E$ is the Clifford module structure on $\E$.
Then $\E_U$ becomes a  $G$-equivariant $\ZZ_2$-graded self-adjoint
Clifford module over $U$. The pair $(\E_U,\v|_U)$ is a tamed
Clifford module. Combining \refc{gluing} with \refl{zerosv}, we
obtain
\eq{M-U}
        \chi_G^{\an}(\E,\v) \ = \ \chi_G^{\an}(\E_U,\v|_U).
\end{equation}

Let us fix a Clifford connection $\n^{\E_U}$ on $\E_U$. It follows
from the proof of \refl{rescaling} (cf. \refs{prrescaling}), that
we can choose an admissible function $f:U\to [0,\infty)$ for the
triple $(\E_U,\v|_U,\n^{\E_U})$ so that $f>1$ and $f(x)\to \infty$
as $x\to\infty$. Then the function
\eq{of}
    \of(x) \ = \
     \begin{cases}
       1/f(x), \quad x\in U;\\
       0, \quad x\not\in U,
     \end{cases}
\end{equation}
is coninuous.

\subsection{A more explicit construction of the topological index}\Label{SS:topind2}
The following explicit construction of $\chi_G^{\top}(\E,\v)$ is
convenient for our purposes.

Let $\rho:[0,\infty)\to [0,\infty)$ be a smooth function such that
$\rho(t)=1$ for $t\le1$ and $\rho(t)=t$ for $t\ge2$. Consider a
new symbol
\eq{sig'}
    \sig'(\xi)
    \ := \
    \frac{\sqrt{-1}}{\rho(|\xi|)}\big(\, c(\xi)+c(v)\,
    \big), \qquad \xi\in T^*M.
\end{equation}
Then $\sig'$ is a symbol of order 0.

Let $U$ be as in \refss{topind}. Then $\sig'(\xi)$ is invertible
for all $\xi\in \pi^{-1}(M\backslash{U})$. We now give a more
explicit than in \refss{indTEcom} construction of the extension of
$\sig'$ to $N$.

Fix an open relatively compact subset $\tilU\subset M$ which
contains the closure of $U$. Then there exists a bundle $F$ over
$\tilU$, such that the bundle $\E^+|_{\tilU}\oplus{F}$ is trivial.
Consider the symbol
\[
    \tilsig' \ := \ \sig'|_{\tilU}\oplus\Id
    \ \in \
    \Gam(\tilU,\Hom(\E^+|_{\tilU}\oplus F,\E^-|_{\tilU}\oplus F).
\]
The map $c(v)\oplus\Id$ defines an isomorphism between the
restrictions of $\E^+|_{\tilU}\oplus{F}$ and
$\E^-|_{\tilU}\oplus{F}$ to $\tilU\backslash{U}$, and, hence, a
trivialization of $\E^-|_{\tilU}\oplus{F}$ over
$\tilU\backslash{U}$.

Let $j:\tilU\to N$ be a $G$-equivariant embedding of $\tilU$ into
a compact $G$-manifold $N$. Then the bundles
$\E^\pm|_{\tilU}\oplus{F}$ extend naturally to bundles $\wE^\pm_N$
over $N$ and the symbol $\tilsig$ extends naturally to a
zeroth-order transversally elliptic symbol $\tilsig'_N$ on $N$,
whose restriction to $N\backslash{U}$ is the identity map.

The symbol $\tilsig'_N$ is homotopic to the symbol $\tilsig_N$ of
\refss{indTEncom}. Hence, these 2 symbols have the same indexes
and we obtain
\eq{indtopE}
    \chi_G^{\top}(\sig) \ := \ \chi_G(\tilsig'_N).
\end{equation}


\subsection{A homotopy of the symbol $\tilsig'_N$}\Label{SS:homot}
Let $\oc:\wE^+\to\wE^-$ denote the map, whose restriction to
$\tilU$ is $c(v)\oplus\Id$ and whose restriction to
$N\backslash{U}$ is the identity map. Recall that the function
$\of$ was defined in the end of \refss{EU}.  Set
\[
    \hatsig_N(\xi) \ = \
        \sqrt{-1}\, \oc \ + \
             \frac{\sqrt{-1}\, \of}{\rho(|\xi|)}c(\xi),
    \qquad \xi\in T^*N.
\]
Clearly, $\hatsig_N$ is homotopic to $\tilsig_N'$.


\subsection{A transversally elliptic operator with symbol $\hatsig_N$}\Label{SS:oper}
We now construct a particular zero-order transversally elliptic
operator $P$ on $N$, whose symbol is equal to $\hatsig_N$ and,
consequently, whose index is equal to $\chi_G^{\top}(\E,\v)$.

Let $A:\Gam(N,\tilE^+)\to \Gam(N,\tilE^+)$ be an invertible
positive-definite self-adjoint $G$-invariant second-order
differential operator, whose symbol is equal to $|\xi|^2$.

Let $D^\pm_U:\Gam(U,\E^\pm_U)\to \Gam(U,\E^\mp_U)$ be the Dirac
operator associated to the Clifford connection $\n^{\E_U}$, cf.
\refss{EU}. Since $\supp\of$ coincides with the closure of $U$, we
can and we will consider the product $\of{D_U}$ as an operator on
$N$.

Set
\[
    P \ = \ \sqrt{-1}\oc \ + \ \of\alp^{-1} D_U^+A^{-1/2}.
\]
Then the symbol of $P$ is equal to $\hatsig_N$
\footnote{
Note, that the symbol of the operator $\alp^{-1}D_U^+$ is equal to
$-\sqrt{-1}c(\xi)$. Therefore, though the function $\alp^{-1}$
tends to infinity near the boundary of $U$, the coefficients of
the differential operator $\of\alp^{-1}D_U^+$ are continuous in
any coordinate chart. Hence, the pseudo-differential operator $P$
is well defined.}.
Hence,
\eq{indP=}
    \chi_G^{\top}(\E,\v) \ = \ \chi_G(P).
\end{equation}

\subsection{A deformation of $P$}\Label{SS:defP}
Consider the family of operators
\[
    P_t \ = \
    (1-t)\sqrt{-1}\, \oc \ + \ t\sqrt{-1}\, \oc A^{-1/2}
        \ + \ \of\alp^{-1} D_U^+A^{-1/2},
    \qquad t\in[0,1].
\]
Then $P_0=P$.

For every irreducible representation $V\in \Irr{G}$, let us denote
by $P_t^V, \ (t\in[0,1])$ the restriction of $P_t$ to the
isotipical component corresponding to $V$.

For each $t_1,t_2$, the difference $P_{t_1}-P_{t_2}$ is a bounded
operator, depending continuously on $t_1$ and $t_2$. Also, for all
$t<1$, the operator $P_t$ is transversally elliptic. Therefore,
for every $V\in \Irr{G}$ and every $t<1$, the operator $P_t^V$ is
Fredholm. Hence, $\chi_G(P_t)= \chi_G(P_0)$ for every $t<1$.
Moreover, to show that $\chi_G(P_1)= \chi_G(P_0)$ we only need to
prove that the operator $P_1^V$ is Fredholm for all $V\in
\Irr{G}$.

\subsection{The operator $P_1$. Proof of \reft{IndTh}}\Label{SS:prIndTh}
Let us investigate $\Ker P_1$. Note, first, that
\[
    P_1 \ = \
    \sqrt{-1}\,  \oc A^{-1/2} \ + \ \of\alp^{-1} D_U^+A^{-1/2}.
\]
Hence, $u\in\Ker{P_1}$ if and only if $w:=A^{-1/2}u$ satisfy
\eq{w=0}
   \big( \sqrt{-1}\, \oc+\of\alp^{-1}D_U^+\big)w=0.
\end{equation}
Since, $\of\equiv0$ and $\oc\equiv\Id$ on $N\backslash{U}$, it
follows from \refe{w=0}, that $w\equiv0$ on $N\backslash{U}$.
Hence, \refe{w=0} is satisfied if and only if $\supp{w}$ lies in
the closure of $U$ and \refe{w=0} holds on $U$. Recall that on $U$
we have $\of=1/f, \ \oc=c(v)$. Hence, \refe{w=0} is equivalent to
\[
    \Big(\, \sqrt{-1}\,  c(v)+\frac{1}{f\alp}D_U^+\, \Big)w
    \ = \ 0
    \ \Leftrightarrow \
    (D_U^++\sqrt{-1}\, f\alp c(v))w \ = \ 0.
\]
Since, $\alp c(v)=c_U(v)$, the later equation is equivalent to
$(D_U^++\sqrt{-1}fc_U(v))w=0$. Since, $A^{-1/2}$ is invertible we
see that $\Ker{P_1}$ is equivariantly isomorphic to
$\Ker(D_U^++\sqrt{-1}fc_U(v))$. Similarly, one shows that
$\Coker{P_1}$ is equivariantly isomorphic to
$\Ker(D_U^-+\sqrt{-1}fc_U(v))$. Therefore
\eq{P1-an}
    \chi_G(P_1) \ = \ \chi_G(D_U+\sqrt{-1}\, fc_U(v))
    \ := \ \chi_G^{\an}(\E|_U,\v|_U).
\end{equation}
In particular, we see that $P_1^V$ is Fredholm for every
$V\in\Irr{G}$. Hence, as it was explained in the end of the
previous subsection,
\[
    \chi_G(P_1) \ = \ \chi_G(P).
\]
\reft{IndTh} follows now from \refe{M-U}, \refe{indP=} and
\refe{P1-an}. \hfill$\square$

\bibliographystyle{amsplain}
\providecommand{\bysame}{\leavevmode\hbox
to3em{\hrulefill}\thinspace}

\end{document}